\documentclass{aa}  

\usepackage{graphicx}
\usepackage{natbib}
\usepackage{upgreek}
\usepackage{tikz}
\graphicspath{{./}{./figures/}}
\usepackage{txfonts}
\usepackage[]{hyperref}

\begin{document} 


\title{Using polarimetry to retrieve the cloud coverage \\
       of Earth-like exoplanets}

\author{L. Rossi\thanks{ \email{l.c.g.rossi@tudelft.nl}}
    \and D. M. Stam
}

\institute{Faculty of Aerospace Engineering, Delft University of Technology,
           Kluyverweg 1, 2629 HS Delft, The Netherlands\\
}

\date{Received February 09, 2017; accepted July 12, 2017}


 
\abstract
   {Clouds have already been detected in exoplanetary atmospheres. They play 
    crucial roles in a planet's atmosphere and climate and can also create 
    ambiguities in the determination of atmospheric parameters such as trace 
    gas mixing ratios. Knowledge of cloud properties is required when assessing the 
    habitability of a planet.}
   {We aim to show that various types of cloud cover such as polar cusps, subsolar clouds, 
    and patchy clouds on Earth-like exoplanets can be distinguished from each other 
    using the polarization and flux of light that is reflected by the planet.}
   {We have computed the flux and polarization of reflected starlight for different 
    types of (liquid water) cloud covers on Earth-like model planets using the 
    adding--doubling method, that fully includes multiple scattering and polarization. 
    Variations in cloud-top altitudes and planet--wide cloud cover percentages 
    were taken into account.}
   {We find that the different types of cloud cover (polar cusps, subsolar clouds,
    and patchy clouds) can be distinguished from each other and that the percentage 
    of cloud cover can be estimated within $10~\%$.}
   {Using our proposed observational strategy, one should be able to determine
    basic orbital parameters of a planet such as orbital inclination and estimate cloud coverage with
    reduced ambiguities from the planet's polarization signals along its orbit.}

   \keywords{polarimetry --
                exoplanets --
                atmospheres --
                clouds
               }

   \maketitle

    \section{Introduction}


    After two decades with huge successes in exoplanet detection, the 
    next step in exoplanetary science is the characterization of planets
    around other stars and the comparison of their properties with those of the
    planets in the solar system. 
    Such a comparison will undoubtedly lead to new understandings and insights
    in the physical processes that form and shape planets, their surfaces,
    atmospheres, climates, and that determine habitability.
    Despite the fast developments in dedicated telescopes and instruments,
    such as the Gemini Planet Imager (GPI) 
    \citep[][]{Macintosh2014} and SPHERE on ESO's VLT \citep[][]{Beuzit2006}, 
    directly observing exoplanets is still an immensely difficult task even for
    the exoplanets closest to us and will remain so for several years, because
    of the low flux of an exoplanet compared to the high flux of the parent
    star that is very close by in angular distance as seen from the Earth.

    A significant contribution to the planetary signal will come from clouds
    in the planetary atmosphere. 
    Clouds can influence a planetary atmosphere and surface in
    several ways \citep[see e.g.,][]{Marley2013}.
    Firstly, clouds scatter and absorb incident light of the parent star and 
    thermal radiation of the planet itself, and with that they
    play a crucial role in the radiative balance of the planet.
    Clouds thus influence the climate, the surface temperature, and in
    particular the presence of liquid surface water. 
    The latter is generally assumed to be essential for the habitability of a planet
    \citep{Kitzmann2010,Yang2013}.  
    Neglecting the presence of clouds in atmospheric modeling
    can lead to underestimating the surface temperature of a
    planet \citep{Kitzmann2011}.
    In order to properly assess the role of clouds in a planetary atmosphere, 
    knowledge about the spatial and temporal distribution of the following cloud 
    properties is needed: the altitude or pressure at the cloud bottom and top, 
    the cloud optical thickness, and the micro-physical properties (i.e., composition, 
    size, shape and thermodynamic phase) of the cloud particles.

    Clouds also change the appearance of spectral features due to gaseous absorption
    that hold information about the chemical composition of the atmosphere of a planet.
    At infrared wavelengths, \citet{Vasquez2013} and \citet{Kitzmann2011} have
    shown that clouds can cause gaseous absorption bands to become
    undetectable in a planet's thermal emission spectrum, depending on their 
    altitude and the vertical distribution of
    the absorbing gas and the atmospheric temperature, for example in the case of 
    the ozone band at $9.6~\upmu$m for planets around type F stars.
    At visible wavelengths, the presence, altitude and horizontal distribution of clouds
    can change the observable depth of absorption bands in spectra of reflected starlight,
    by scattering light back to space before it reaches the absorber, and/or by increasing
    the average optical path through the atmosphere \citep{Fauchez2017}.
    Clouds can also hide biosignatures from the surface, in particular the 
    so--called red-edge, the steep increase in the albedo of vegetation 
    between the visible and the near infrared
    \citep{Tinetti2006b,Montanes-Rodriguez2006,Seager2005}.
    And finally, in transit observations, \citet{Line2016} showed that when analyzing 
    the stellar spectrum that is filtered through the upper layers of a planetary atmosphere
    during a planetary transit, clouds along the limb will not only influence the
    retrieved amount of absorbing gas, by blocking stellar light, but their influence on 
    the measured spectrum can also mimic the signals of a high mean molecular mass of the
    atmosphere.

    In this article, we investigate the influence of cloud properties on the
    degree and direction of polarization of starlight that is reflected by
    a planet, focusing on the influence of cloud-top pressure, the cloud coverage 
    fraction, and the spatial distribution of the clouds across the planet.
    Polarimetry promises to be a very powerful method in the detection and
    especially in the characterization of exoplanets.
    Polarimetry can be used to detect exoplanets because integrated across 
    their disk, the light of solar type stars can be considered to be 
    unpolarized \citep{Kemp1987} while the starlight that has been reflected 
    by a planet will generally be polarized. 
    Polarimetry thus enhances the contrast between a star and its planet.
    
    The degree and direction of polarization of a planet depend not only on the 
    illumination and viewing directions, and thus on the planet's phase angle
    (the angle between the observer and star measured from the center of the
    planet), but also on the
    composition and structure of the planetary atmosphere and surface (if present)
    \citep{Seager2000,Stam2004}, and measurements of the degree and direction of 
    polarization can be used to retrieve the atmospheric and surface properties.
    A famous example of this use of polarimetry is the derivation of the 
    composition and size of the particles constituting Venus's upper clouds from
    Earth--based observations of the planet's disk--integrated polarization 
    at a few wavelengths and across a wide phase angle range \citep[][]{Hansen1974a}.
    From Earth, the solar system's outer planets can only be observed at a narrow phase
    angle range around 0$^\circ$, where the degree of polarization is usually very 
    small, because mainly backscattered light is observed 
    \citep[see e.g.,][and references therein]{McLean2017}.
    Exoplanets will usually be observable at a large phase angle range (except if
    the planetary orbit is seen face-on, thus with an inclination close to 0$^\circ$: 
    then the phase angle will always be around 90$^\circ$).
    An exoplanet's polarization signal will thus usually vary as the planet 
    orbits its star.
    
    The structure of this article is as follows.
    In Sect.~\ref{sec:numerical}, we introduce the algorithms we use for the radiative 
    transfer computations
    and the integration of flux and polarization signals across the visible and
    illuminated part of a planetary disk. 
    In Sect.~\ref{sec:cloud-covers}, we describe the model atmospheres and 
    the cloud properties for our model planets.
    In Sect.~\ref{sec:results}, we present our numerical results and 
    compare the influence of different types of cloud covers on the
    flux and polarization signals of reflected light from a planet across all
    phase angles, including a discussion on ambiguities that
    can arise when retrieving cloud properties from measured signals.
    Finally, in Sect.~\ref{sec:obs-strat}, we discuss an observational strategy 
    that could be used to derive cloud coverage with reduced ambiguities,
    and in Sect.~\ref{sec:conclusion}, we present our conclusions. 


    \section{Numerical algorithms}
    \label{sec:numerical}


    \subsection{Defining fluxes and polarization}

    We describe the starlight that is incident on a planet and the starlight
    that is reflected by the planet by Stokes vectors, as follows
    \citep[see, e.g.,][]{Hansen1974,Hovenier2004}
    \begin{equation}
        \mathbf{F} = \left[ \begin{array}{c}
                F \\ Q \\ U \\ V
        \end{array} \right],
        \label{eq:def-stokes-vector}
    \end{equation}
    where $F$ is the total flux, $Q$ and $U$ are the linearly polarized fluxes and
    $V$ is the circularly polarized flux. These fluxes are usually expressed in 
    W~m$^{-2}$ or, for example, in W m$^{-2}$ nm$^{-1}$ when used spectrally resolved. 

    We have assumed the starlight that is incident on a planet to be unpolarized
    \citep[see][]{Kemp1987}. This incident light will be described by 
    ${\bf F}_0 = F_0 \bf{1}$, with $\pi F_0$ the stellar flux measured
    perpendicular to the direction of propagation, and ${\bf 1}$ the
    unit column vector. 
    Starlight that is reflected by an orbiting planet will usually be polarized.
    Stokes parameters $Q$ and $U$ of this light are defined with respect to a
    reference plane, for which we use the planetary scattering plane, that is,\ the
    plane that contains the star, the planet and the observer. 
    We have ignored the circularly polarized flux $V$ as its values are very small 
    compared to $Q$ and $U$, and ignoring $V$ causes no significant errors in the 
    computed values of $F$, $Q$, and $U$ \citep{Stam2005}.

    The degree of linear polarization is defined as 
    \begin{equation}
        P_\ell = \frac{\sqrt{Q^2 + U^2}}{F}.
        \label{eq:def-pol}
    \end{equation}
    For a planet that is symmetric with respect to the planetary scattering plane, 
    the disk--integrated flux $U$ will be zero, and the degree of linear 
    polarization can then be defined as 
    \begin{equation}
        P_{\rm s} = \frac{-Q}{F}.
        \label{eq:def-pol-q}
    \end{equation}
    This definition includes the direction of polarization:
    $P_{\rm s}$ is positive if the polarization is perpendicular to
    the planetary scattering plane and negative if it is parallel to the
    plane.

    The degree of polarization that we present in this paper pertains
    to a planet that is observed spatially resolved from its parent star, 
    without any background starlight. In actual observations, even a spatially
    resolved planet will be surrounded by some background starlight,
    depending on for example, the distance between the planet and its star,
    the brightness of the star, the distance between the observer and the
    exoplanetary system, and the telescope and the instrument capabilities,
    such as coronagraphs and/or adaptive optics.
    The amount of background light will also depend on the wavelength.
    In the presence of background starlight, the observable degree of
    polarization in the detector pixel that contains the planet
    would be given by
    \begin{equation}
        P_{\ell *} = \frac{\sqrt{Q^2 + U^2}}{F + F_*},
    \end{equation}
    with $F_*$ the background stellar flux (that is assumed to be unpolarized)
    in the pixel.  
    \citet{Seager2000} show simulations of $P_{\ell *}$ of spatially unresolved
    hot Jupiters where $F_*$ is the full stellar flux. Because of the huge
    difference in $F$ and $F_*$, $P_{\ell *}$ is then of the order of 10$^{-6}$,
    depending on the size of the planet, its atmospheric composition and the
    phase angle. 
    Detections of such spatially unresolved exoplanets in polarimetry have not 
    yet been confirmed, but first attempts 
    seem promising \citep[][and references therein]{Bott2016}.
    Instruments like EPICS on the E-ELT \citep{Kasper2010,Keller2010} will combine 
    coronagraphs and extreme adaptive optics to limit $F_*$ and increase $P_{\ell *}$.
    In the following, we assume $F_{*}= 0$. The polarization values in
    our results should thus be regarded as upper limits.
    

    \subsection{Orbital parameters}

    All the computations in this study were conducted under the assumption that
    the model planet is in an `edge-on'-orbit as seen by the observer (the
    inclination angle $i$ of the orbit is thus $90^\circ$). 
    The planet itself is assumed to be spherical and with no obliquity, 
    so the sub-observer point was always located on the equator of the planet.
    As the planet's orbital plane was assumed to be aligned with the horizontal axis in
    the observer's reference frame, the spin axis of the planet is
    vertical as seen by the observer.

    \subsection{The radiative transfer algorithm}

    While our model planets are spatially resolved from their star, they themselves
    are spatially unresolved, meaning that\ each planet is observed as a single speck of light.
    We computed the disk--integrated Stokes parameters and polarization $P_{\ell}$ of
    a model planet with the following four~steps. \\

    \noindent
    {\bf Step 1.} We projected the planetary disk as seen by the observer on the
    plane of the sky and divide the disk--circumscribing square into an
    $n_{\rm pix} \times n_{\rm pix}$ grid of square, equally sized pixels with
    the planet's equator aligned with the (horizontal) $x-$axis.
    We projected the center of each pixel onto the spherical planet (discarding 
    pixels with their centers outside the planetary disk) to identify the
    location on the planet for which we will compute the locally reflected 
    Stokes vector.
    Increasing $n_{\rm pix}$ (i.e.,\ the spatial resolution on the planet)
    increases the accuracy of our computations, but it also increases the 
    computation time.
    
    The number of pixels required to reach a given accuracy increases
    with increasing phase angle $\alpha$ (the angle between the star and the observer
    as measured from the center of the planet). 
    Indeed, when the planet is close to `full', the value for $n_{\rm pix}$
    required for an accurate result is much smaller than when the signal comes
    from a narrow crescent of the planet. 
    As a compromise between sufficient resolution and acceptable computation
    time, we have used an adaptable value for $n_{\rm pix}$, given by the
    following equation:
    \begin{equation}
       n_{\rm pix}(\alpha) = n_{\rm pix}(0^\circ) \left[ 1 + \sin^2 (\alpha/2) \right],
    \label{eq:def-npix}
    \end{equation}
    with $n_{\rm pix}$ rounded up to the nearest integer. 
    Unless stated otherwise, $n_{\rm pix}(0^\circ)= 40$, and
    thus $n_{\rm pix}(180^\circ)=80$ (see Appendix~\ref{app:npix} for a 
    discussion on optimizing $n_{\rm pix}$).
    With this pixel approach, we could straightforwardly model horizontally
    inhomogeneous planets by choosing a different atmosphere and/or surface model
    for different pixels. These models are described in Sect.~\ref{sec:cloud-covers}. \\
    
    \noindent
    {\bf Step 2.} For each projected pixel on the planet, we determined
    the following angles: $\theta_0$, the angle between the local zenith
    direction and the direction toward the star, $\theta$, the angle between
    the local zenith direction and the direction toward the observer, and the
    azimuthal difference angle $\phi-\phi_0$, the angle between the
    plane containing the local zenith direction and the direction toward the
    star and the plane containing the local zenith direction and the direction
    toward the observer \citep[see][]{deHaan1987}.
    The angles depend on the latitude and longitude of the projected location 
    on the planet (the sub--observer longitude and latitude equal 0$^\circ$), and
    $\theta_0$ and $\phi-\phi_0$ also depend on phase angle $\alpha$.
    We note that we used the azimuthal difference angle rather than $\phi$ and
    $\phi_0$ separately because the planetary atmosphere and surface of each
    pixel are rotationally symmetric with respect to the local zenith
    direction. Pixels with $\theta_0 > 90^\circ$ are assumed to be black, as 
    the parent star is below the local horizon. \\

    \noindent
    {\bf Step 3.} For each projected pixel on the planet
    and the local atmosphere-surface model, we calculated 
    the Stokes vector of the reflected starlight according to
    \citep[see, e.g.,][]{Hansen1974}
    \begin{equation}
        {\bf F}(\theta,\theta_0,\phi-\phi_0)= \cos{\theta_0} \hspace{0.2cm}
        {\bf R}_1(\theta,\theta_0,\phi-\phi_0) \hspace{0.2cm}
        {\bf F}_0,
        \label{eq:reflected_vector}
    \end{equation}
    with ${\bf R}_1$ the first column of the $3 \times 3$ local planetary reflection matrix
    (we ignore the other columns, since the incoming starlight is unpolarized). 
    We computed ${\bf R}_1$ with an adding--doubling algorithm that fully includes
    polarization for all orders of scattering \citep[based on][]{deHaan1987}.
    Rather than embarking on a separate radiative transfer computation for every 
    pixel, we first computed and store the coefficients ${\bf R}_1^m(\theta_0,\theta)$
    of the expansion of ${\bf R}_1(\theta_0,\theta,\phi-\phi_0)$ into a Fourier series 
    ($0 \leq m < M$, with $M$ the total number of coefficients)
    for the different atmosphere--surface models (typically two) on the
    model planet. 
    Our adding--doubling algorithm computes these coefficients at values of
    $\cos \theta_0$ and $\cos \theta$ that coincide with Gaussian abscissae,
    the total number of which is user--defined. 
    For increased accuracy in the disk--integration, we also computed
    coefficients at $\cos \theta_0= 1$ ($\theta_0=0^\circ$) and $\cos \theta=
    1$ ($\theta= 0^\circ)$.
    Given a pixel with local values of $\theta_0$, $\theta$, and $\phi-\phi_0$,
    we were able to efficiently compute its ${\bf R}_1$ by summing up
    the Fourier coefficients stored for the
    appropriate atmosphere--surface model, interpolating when
    necessary. \\
    
    \noindent
    {\bf Step 4.} 
    A locally reflected Stokes vector as computed using our adding-doubling
    algorithm is defined with respect to the local meridian plane, which
    contains both the local zenith direction and the direction toward the
    observer. 
    We had to redefine each locally defined vector to the common reference
    plane, that is, the planetary scattering plane, with
    a rotation matrix \citep[see][]{Hovenier1983}
    and the local rotation angle measured between the local meridian plane and
    the planetary scattering plane (for details, see Appendix~\ref{app:rot}).
    Then we computed the disk--integrated Stokes vector by summing up the local, 
    redefined Stokes vectors.
    The actual area on the three dimensional planet that is covered by a
    projected square pixel varies with the latitude and longitude, 
    but because all square pixels have the same size, their respective
    Stokes vectors as calculated by Eq.~\ref{eq:reflected_vector} contribute
    equally to the disk--integrated planetary signal.
    
    We finally normalized each disk--integrated Stokes vector such that at 
    $\alpha=0^\circ$, flux $F$ equals the planet's geometric albedo.
    The degree of polarization that we computed from the disk--integrated
    Stokes vector was independent of this normalization
    because it is a relative measure (see Eqs.~\ref{eq:def-pol} 
    and~\ref{eq:def-pol-q}).


    \section{Atmosphere and surface models}
    \label{sec:cloud-covers}
    
    Locally, the atmospheres of our model planets are composed of stacks of
    horizontally homogeneous layers, filled with gas and, optionally, cloud particles,
    above a flat, Lambertian (i.e., isotropic and unpolarized) reflecting surface 
    with albedo $a_{\rm surf}$.
    We assumed an Earth--like gas mixture in each layer, with a depolarization 
    factor $\delta=0.03$ and a molecular mass of $29$~g/mol. We do not consider
    absorption by the gas. 
    Table~\ref{tab:params} lists the atmospheric parameters. 
    Our cloud models are described in more detail below.


    \subsection{Physical properties of the clouds}

    We used liquid water clouds. The refractive index of the cloud particles
    is $n_r=1.33+10^{-8}i$ \citep{Hale1973}.
    The particle size distribution is a two--parameter gamma distribution
    \citep[][]{Hansen1974a} with $r_{\rm eff} = 8.0~\upmu$m and $\nu_{\rm eff} = 0.1$,
    based on Earth cloud values from \citet{Han1994}.
    All clouds have an optical thickness of 6.0 \citep{Warren2007}, independent
    of the wavelength.
    We have not investigated the effect of varying the particle size distribution 
    and/or the optical thickness of the clouds as this has been studied by
    \citet{Karalidi2011, Karalidi2012}.
    The clouds span a vertical extent of $100$~mb and we set their
    altitude to represent low- to mid-altitude clouds corresponding to
    cumulus, stratus and stratocumulus \citep[see][]{Rossow1999,Hahn2001}.
    Table~\ref{tab:params} includes the cloud parameters.

    \begin{table}
        \centering
        \begin{tabular}[h]{l l c}
            Parameter & Symbol & Value \\ \hline
            Surface albedo & $a_{\rm surf}$ & 0.0 \\
            Surface pressure [bar] & $p_{\rm surf}$ & 1.0 \\
            Depolarization factor & $\delta$ &  0.03 \\
            Mean molecular mass [g/mol] & $m_{\rm g}$ & 29 \\
            Acceleration of gravity [m/s$^2$] & $g$ & 9.81 \\
            Cloud particle effective radius [$\upmu$m] & $r_{\rm eff}$ & 8.0 \\
            Cloud particle effective variance & $\nu_{\rm eff}$ & 0.1 \\
            Cloud optical thickness & $\tau_{\rm c}$ & 6.0 \\
            Cloud-top pressure [bar] & $p_{\rm c}$ & 0.6; 0.7; 0.8 \\
            \hline
        \end{tabular}
        \caption{Parameters of our standard model atmosphere and surface.}
        \label{tab:params}
    \end{table}

    \subsection{Cloud covers}


    We investigated three different types of cloud coverages: sub-solar clouds,
    polar--cusps, and patchy clouds. 
    We modeled these clouds by assigning specific pixels to be cloudy. 
    All other pixels on the planet are cloud--free. \\
    
\begin{figure}[h]
        \centering
        \begin{tabular}{c}
            \includegraphics[width=0.4\linewidth]{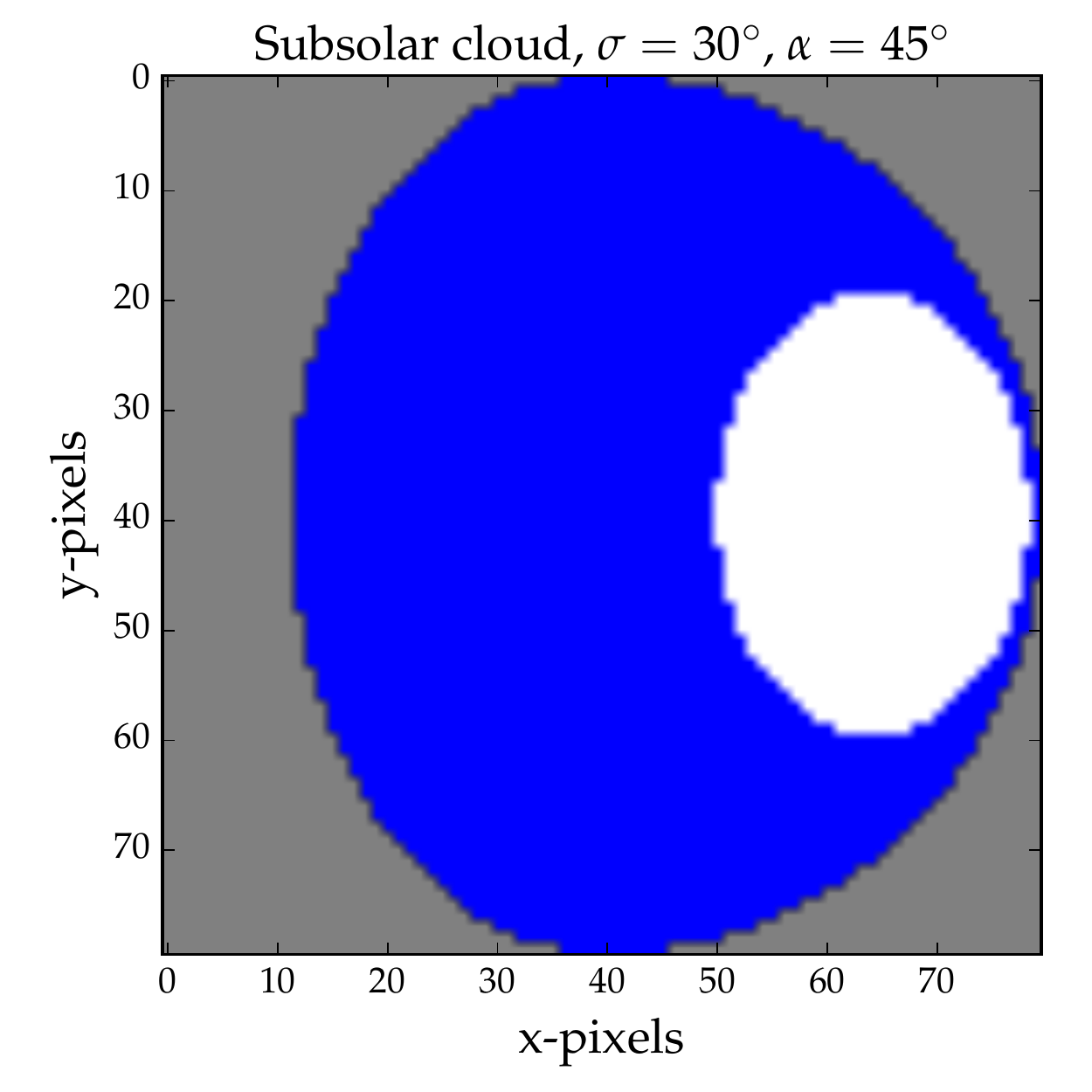} \\
            \includegraphics[width=0.4\linewidth]{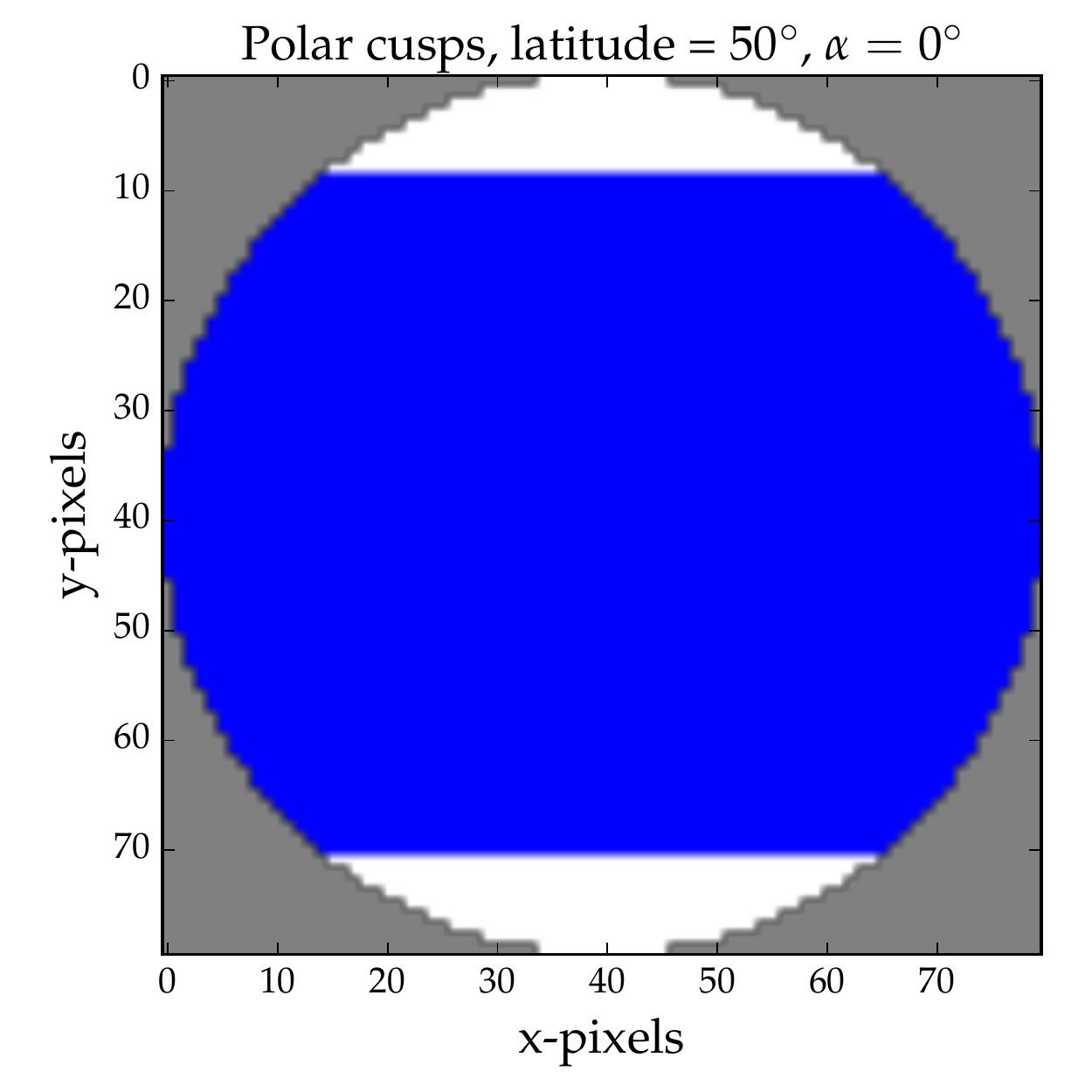} \\
            \includegraphics[width=0.4\linewidth]{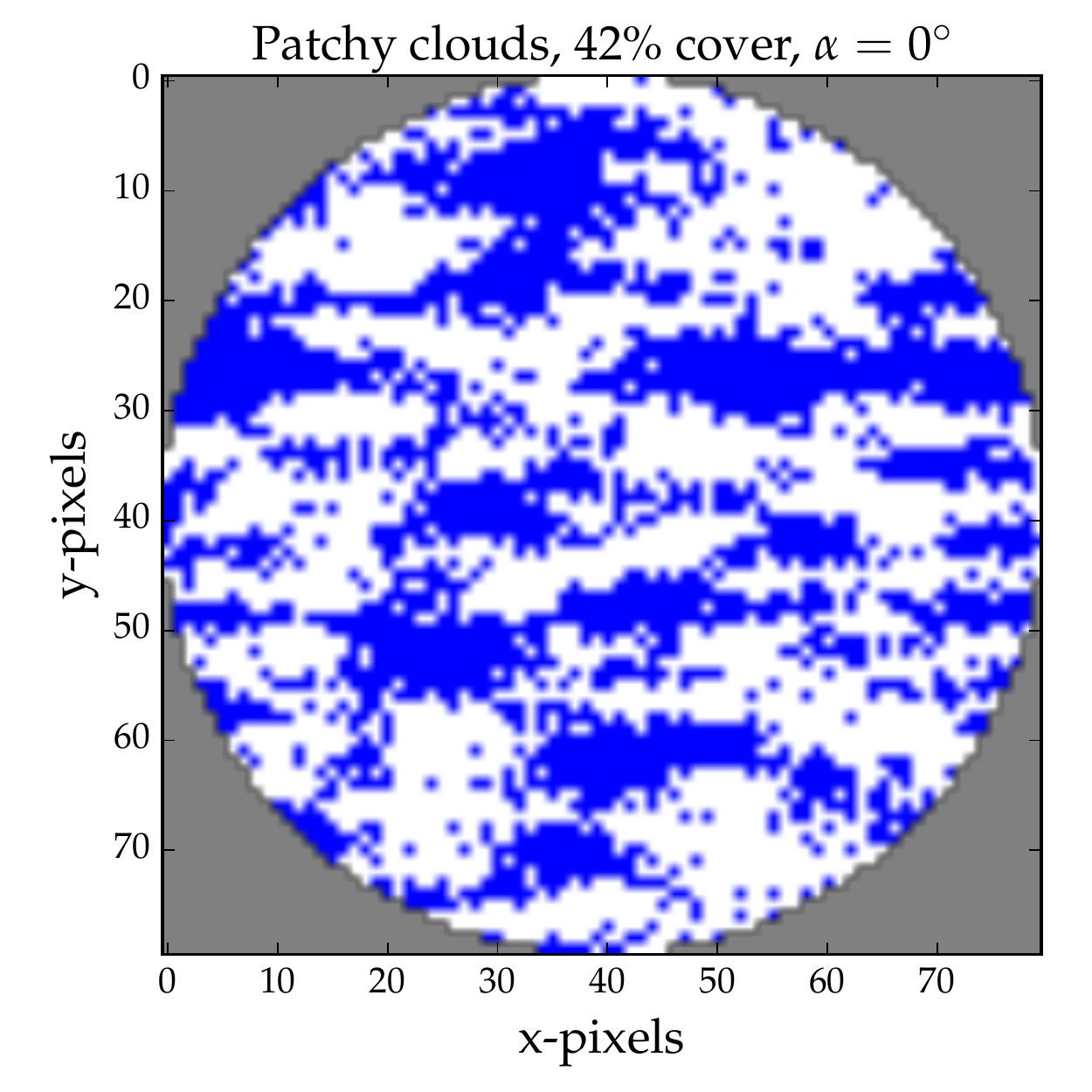}
        \end{tabular}
        \caption{Examples of our three types of cloud cover on a
            $80 \times 80$ pixel grid:
            (a) Sub-solar clouds for $\sigma_{\rm c}=30^\circ$
            and $\alpha=45^\circ$;
            (b) Polar cusps for $L_{\rm t}=50^\circ$
            and $\alpha=0^\circ$;
            (c) Patchy clouds for $F_{\rm c}=0.42$ and $\alpha=0^\circ$.}
        \label{fig:cover_types}
    \end{figure}

    \noindent
    {\em Sub-solar clouds} are relevant for tidally--locked exoplanets \citep{Yang2013}. 
    To model these clouds, the pixel grid was filled such that only the region on a planet
    with the local solar zenith angle $\theta_0$ smaller than a given angle 
    $\sigma_{\rm c}$ is cloudy (see Fig.~\ref{fig:cover_types}a). \\

    \noindent
    {\em Polar--cusps} are clouds that form where the daily averaged incident stellar flux
    is below a certain threshold. 
    In this model, the cloudy pixels are located above a threshold latitude
    $L_{\rm t}$ on the planet (see Fig.~\ref{fig:cover_types}b). \\

    \noindent
    {\em Patchy clouds} can be anywhere on the planet. 
    They are described by $F_{\rm c}$, the fraction of all pixels on the whole
    disk that are cloudy, and the actual spatial distribution of cloudy pixels
    across the planet (see Fig.~\ref{fig:cover_types}c).
    We generated patchy clouds by drawing 50 values from a 2D Gaussian distribution
    centered on a location randomly chosen within the $n_{\rm pix} \times
    n_{\rm pix}$ grid. 
    The covariance matrix is given by 
    \begin{equation}
        \Sigma = n_{\rm pix}
        \left[
            \begin{array}{c c}
                x_{\rm scale} & 0 \\
                0 & y_{\rm scale} \\
            \end{array}
        \right]
    ,\end{equation}
    where $x_{\rm scale}$ and $y_{\rm scale}$ are used to
    fine--tune the shape of the cloud patches along the north--south and east--west 
    axes.
    We used $x_{\rm scale}= 0.1$ and $y_{\rm scale}= 0.01$ as nominal values in 
    order to generate clouds with a streaky, zonal--oriented
    pattern similar to that observed on Earth. 
    Cloud patches are generated across the planetary disk until the desired 
    $F_{\rm c}$ is reached.
    We defined $F_{\rm c}$ at $\alpha=0^\circ$, because
    the planetary--wide cloud coverage is more relevant in terms of climatology
    than the coverage seen by the observer. 
    The actual cloud fraction observed at a given angle $\alpha$
    larger than 0$^\circ$ can thus differ from the specified value of $F_{\rm c}$.

    \section{Polarization signatures of different cloud covers}
    \label{sec:results}


    In this section, we compare the disk--integrated polarization 
    of starlight reflected by our model planets for the different
    types of cloud cover defined above.

    \begin{table}[b]
        \centering 
        \begin{tabular}{c c c c c}
            \hline
            Patchy clouds & \multicolumn{2}{c}{Sub-solar clouds} & 
            \multicolumn{2}{c}{Polar cusps} \\
            $F_{\rm c}$ & $\sigma_{\rm c}$ ($^\circ$) & $F_{\rm eff}$ & 
            $L_{\rm t}$ ($^\circ$) & $F_{\rm eff}$ \\
            \hline
            0.10 & 20 & 0.12 & 50 & 0.14 \\
            0.40 & 40 & 0.41 & 30 & 0.39 \\
            0.60 & 50 & 0.59 & 20 & 0.58 \\
            0.80 & 60 & 0.75 & 10 & 0.77 \\
            \hline
        \end{tabular}
        \caption{Cloud covers of equivalent cloud fraction with the parameters used to generate them. 
        $F_{\rm c}$ is the fraction of the planet covered by patchy clouds, 
        $\sigma_{\rm c}$ is the angular width of the sub-solar clouds, 
        $L_{\rm t}$ is the threshold latitude of polar cusps and
        $F_{\rm eff}$ is the effective coverage i.e., the actual coverage for the
        considered distribution of sub-solar clouds and polar cusps.}
        \label{tab:eq-cloud-covers}
    \end{table}

    \subsection{Sub-solar clouds and polar cusps}
 
\begin{figure}[h]
        \centering
        \includegraphics[width=0.9\linewidth]{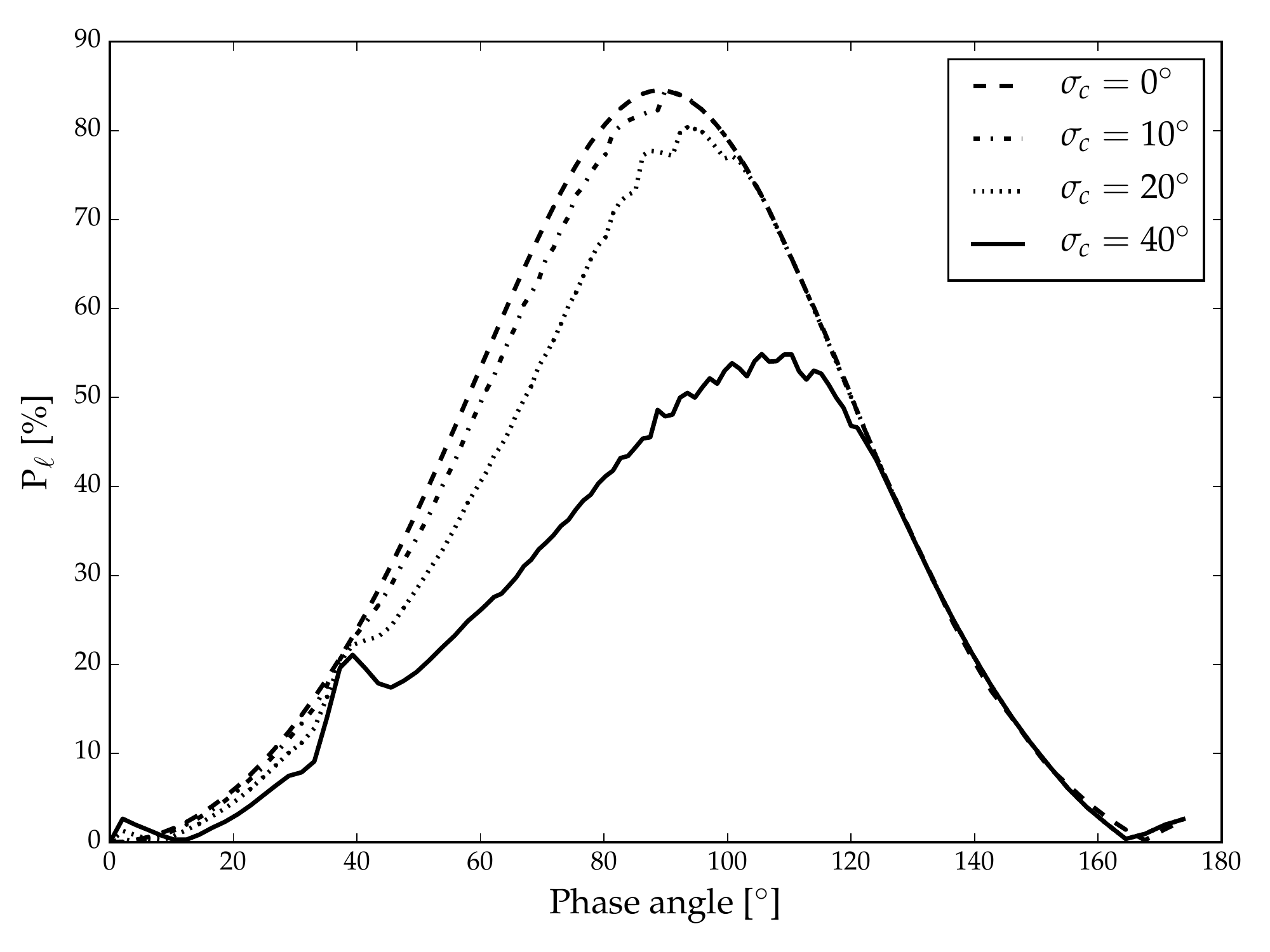}\\
        \caption{Degree of linear polarization $P_\ell$ at $\lambda=500$~nm as a
            function of phase angle $\alpha$ for a sub-solar cloud with $p_{\rm c}=800$~mb,
            for different values of $\sigma_c$. For $\sigma_{\rm c}=0^\circ$,
            the planet is cloud--free. The angular feature around $\alpha=40^\circ$
            is the (primary) rainbow.
            }
        \label{fig:compare-ss}
    \end{figure}

    Figure \ref{fig:compare-ss} shows the degree of linear polarization $P_\ell$ at
    500~nm for different angular sizes $\sigma_{\rm c}$ of the sub-solar cloud
    as a function of $\alpha$ (recall that the actual range of phase angles
    that an exoplanet can be observed at, depends on the orbital inclination
    angle $i$; this is the largest range occurring along the orbit with $i=90^\circ$). 
    The relation between the values of $\sigma_{\rm c}$ that are used 
    and the effective cloud coverage $F_{\rm eff}$
    are given in Table~\ref{tab:eq-cloud-covers}.

    As expected, different values of $\sigma_{\rm c}$, and thus different
    cloud fractions, yield different curves with common features. 
    First, as $\sigma_{\rm c}$ increases, the (primary) rainbow feature near 
    $\alpha=40^\circ$ that is due to
    light scattered by spherical water cloud droplets  
    \citep[][]{Karalidi2012,Bailey2007} becomes more distinct (its maximum
    value decreases slightly).
    The angular location of the rainbow is determined by the micro--physical
    properties, mostly the refractive index, of the clouds particles \citep[see
    e.g.,][]{Hansen1974}. The small bump in $P_\ell$ at phase angles below 10$^\circ$
    is due to the glory that arises from light that is backscattered by the spherical 
    cloud particles \citep[see][]{Hansen1974}.
    
    Furthermore, while the different model planets show different values of
    $P_\ell$ at small $\alpha$ (except at $\alpha=0^\circ$, where $P_{\rm \ell}=0$ 
    for each of these planets, because they are mirror--symmetric with respect 
    to the reference plane), their values of $P_\ell$ are very
    similar at large phase angles. 
    In fact, $P_\ell$ of these planets is dominated by that of the
    cloud as long as it is on the illuminated and visible part of the planet.
    As the tidally locked planets move along their orbit, the clouds disappear
    from the observer's view at a phase angle that depends on $\sigma_{\rm c}$
    and on a possible offset of the cloud with respect to the sub-solar point
    (not shown in Fig.~\ref{fig:compare-ss})
    \citep[as seems to be the case for Kepler 7b,][]{GarciaMunoz2015a}, leaving
    only the $P_\ell$ due to the gas. The maximum polarization due to the Rayleigh
    scattering gas is very
    high (around $\alpha=90^\circ$) because at this wavelength, there is 
    little multiple scattering and the surface is black.
    We note that the oscillations of $P_\ell$ just before the clouds disappears
    completely from view are due to the pixellation of the cloud, which becomes
    more apparent when the visible part of the cloud narrows while 
    the cloud is disappearing across the limb of the planet. These
    oscillations decrease when the number of pixels is increased. 

\begin{figure}[h]
        \centering
        \includegraphics[width=0.9\linewidth]{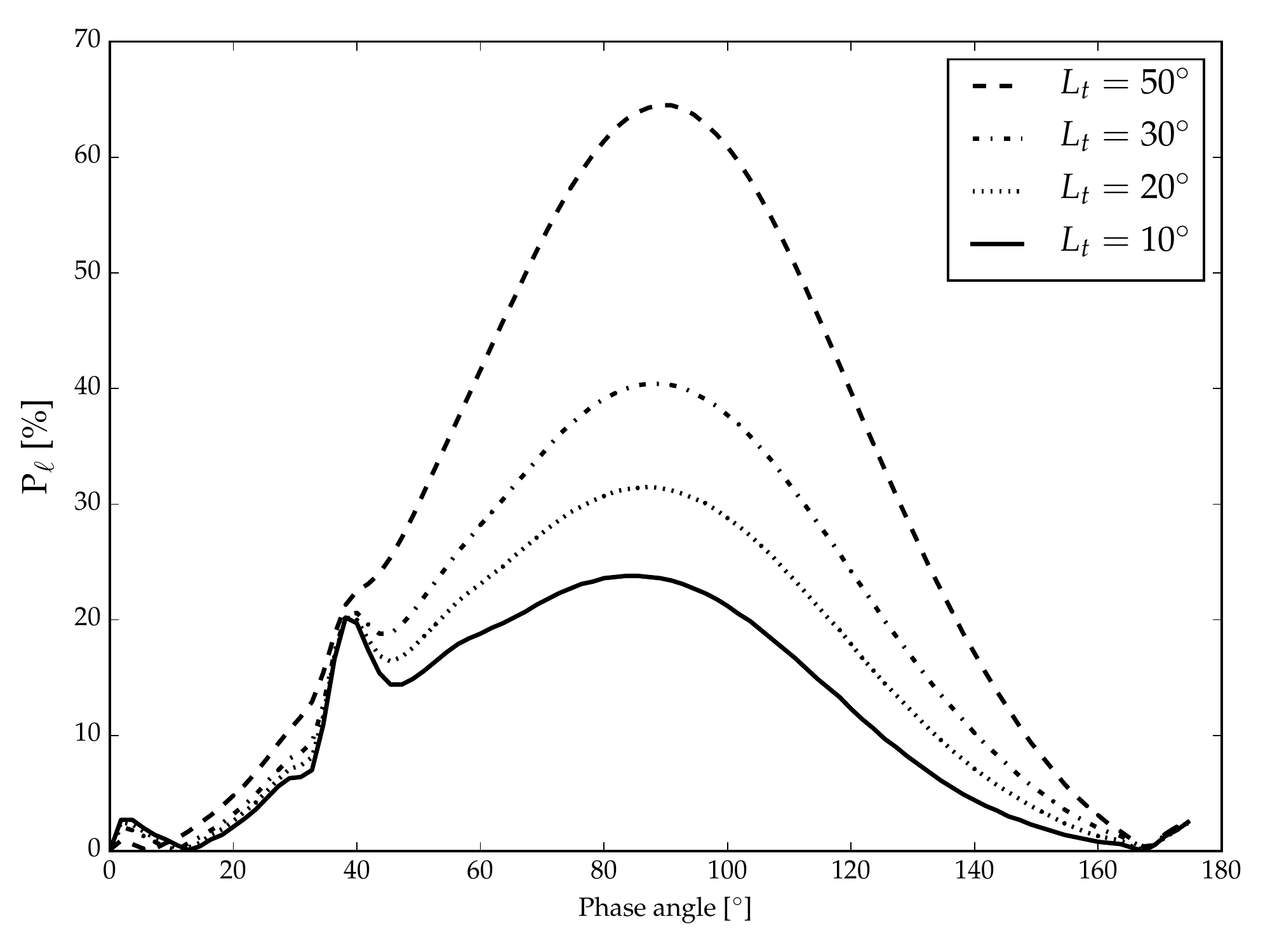}
        \caption{Similar to Fig.~\ref{fig:compare-ss} except for 
        polar cusps clouds for different values of
        the threshold latitude $L_{\rm t}$.}
        \label{fig:compare-cusps}
    \end{figure}

    Figure \ref{fig:compare-cusps} is similar to Fig.~\ref{fig:compare-ss},
    except for polar cusps for different values of the threshold latitude $L_{\rm t}$.
    (the relation between $L_{\rm t}$ and the effective clouds coverage $F_{\rm eff}$ 
    is given in Table~\ref{tab:eq-cloud-covers}).
    Polar cusps clouds exhibit a more
    continuous behavior of the polarization than sub--solar clouds, because they
    remain in view as our model planets rotate. 
    Like with the sub-solar clouds, the rainbow feature is clearly visible near
    $\alpha=40^\circ$. The peak of $P_\ell$ around $\alpha=90^\circ$,
    is again due to Rayleigh scattering.
    The smaller $L_{\rm t}$, thus the larger $F_{\rm eff}$, the stronger 
    the rainbow and the lower the peak of polarization near 90$^\circ$, because 
    the smaller the contribution of (highly polarized) Rayleigh scattered light.
    
    \subsection{Patchy clouds}

    Patchy clouds are interesting because each pixel on the planet has its
    specific illumination and viewing geometries (even though phase angle
    $\alpha$ is the same for all pixels) and therefore contributes its own
    polarization signal to the disk--integrated signal. 
    The precise locations of the cloudy pixels on the disk thus influence
    $P_\ell$ of the planet, and can give rise to different $P_\ell$
    values for the same cloud coverage fraction $F_{\rm c}$. 
    Because of this, for a given value of $F_{\rm c}$ and for each phase angle
    considered, we generate 300 independent, random cloud patterns. 
    The curves shown in subsequent figures for a given $F_{\rm c}$ are the
    averages of these 300 patterns. 
    This allows us to explore the range of possible values of the disk--integrated
    polarization due to different locations of the cloud patches on the planet. 
    This variability is not directly related to temporal variations, because
    the 300 patterns are independent: they do not depend on the rotation of
    the planet, the position of the patches is purely random and not bound
    to a realistic climate model.

\begin{figure}[h]
        \centering
        \includegraphics[width=.9\linewidth]{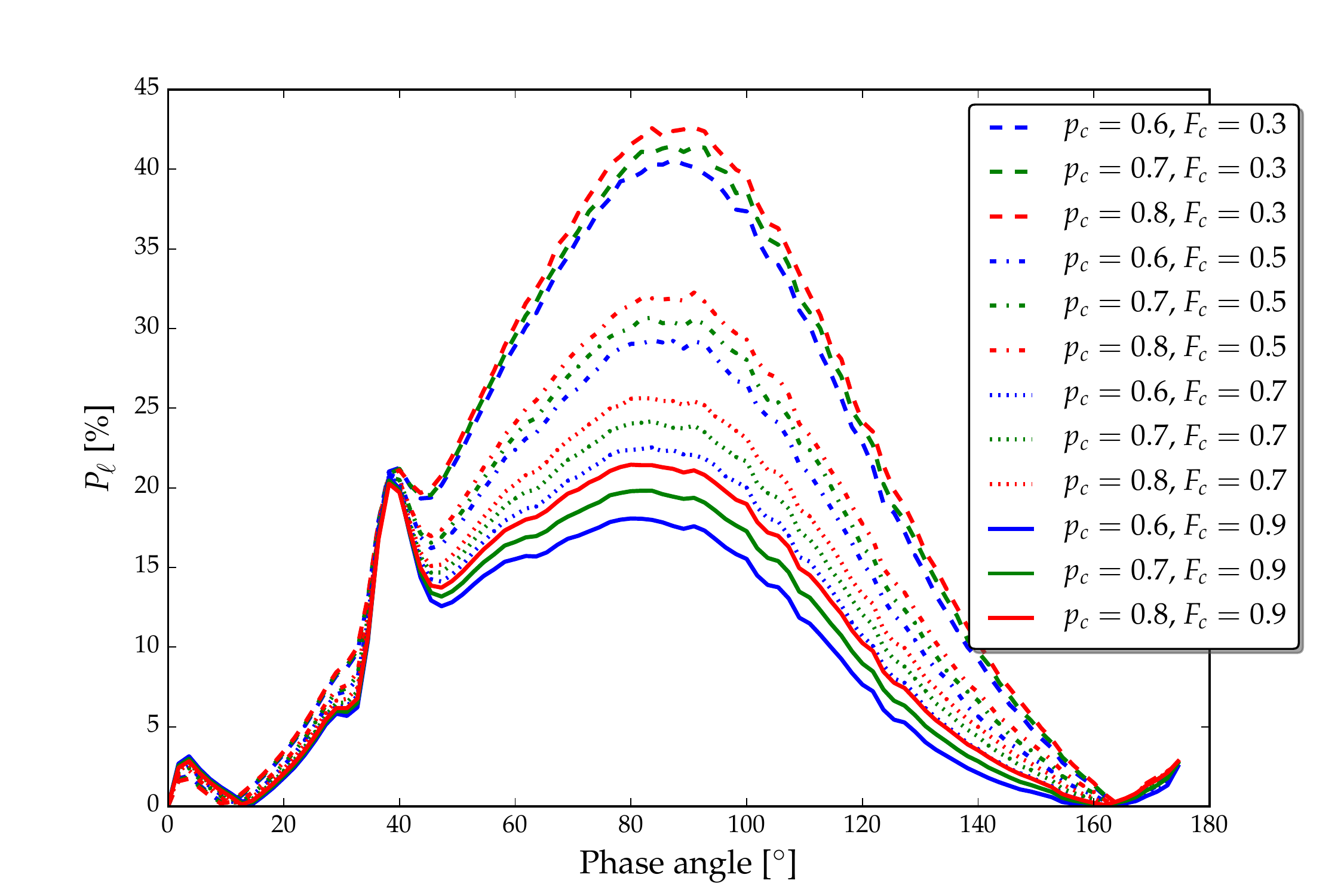}
        \caption{$P_\ell$ at $\lambda=500$~nm 
            for different cloud-top pressures $p_{\rm c}$ (600, 700, and
            800~mb) and cloud coverages $F_{\rm c}$ (0.3, 0.5, 0.7, and 0.9) for
            planets with patchy clouds. Different line--styles indicate different
            values of $F_{\rm c}$ and different colors indicate different values of 
            $p_{\rm c}$. Each curve is the average of 300 curves pertaining to
            300 randomly generated patchy cloud patterns for the given values
            of $F_{\rm c}$ and $p_{\rm c}$.}
        \label{fig:pc_fc}
    \end{figure}

    Figure~\ref{fig:pc_fc} shows $P_\ell$ for different combinations of
    $F_{\rm c}$ and cloud-top pressures $p_{\rm c}$.
    The change of the strength of the rainbow and the peak around $\alpha=90^\circ$
    is similar to what was seen in Fig.~\ref{fig:compare-cusps}.
    For a given value of $F_{\rm c}$, a larger value of $p_{\rm c}$
    is related to smaller values of $P_\ell$, in particular around $\alpha=90^\circ$.
    This is due to the different amount of gas above the clouds: a larger cloud-top pressure (i.e., a lower cloud-top altitude) leaves
    more gas above the clouds and thus more relatively highly polarized 
    Rayleigh scattered light (with increasing gas optical thickness above the clouds,
    the polarization would reach a maximum value before starting to decrease
    due to the increase of multiple scattering).

\begin{figure}[h]
        \centering
        \includegraphics[width=.9\linewidth]{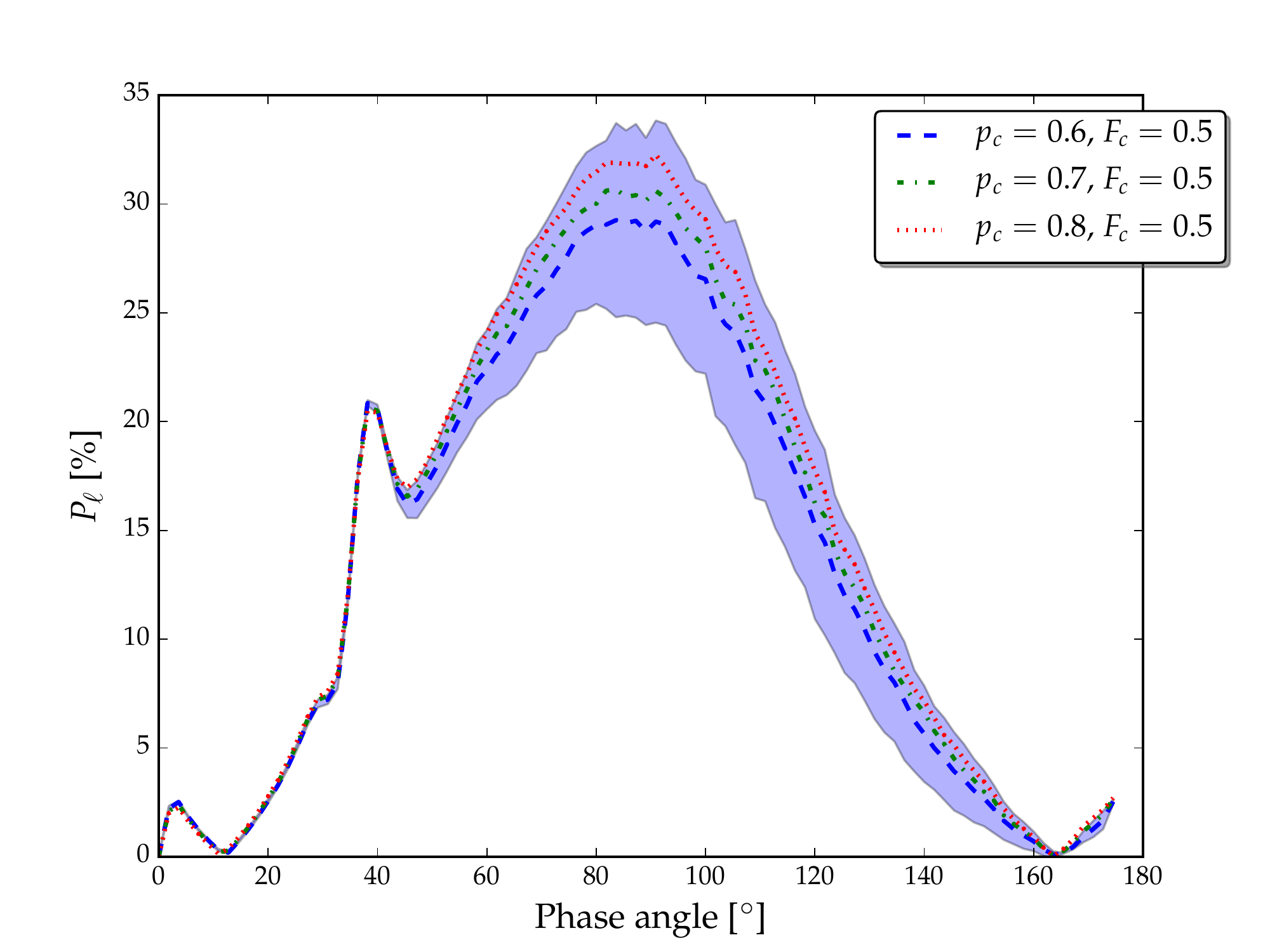}
        \caption{Similar to Fig.~\ref{fig:pc_fc}, except only for $F_{\rm c}=0.5$.
                 The shaded area indicates the 2$\sigma$ variability for the
                 $p_{\rm c}=600$~mb curve.}
        \label{fig:pressure_compare}
    \end{figure}

    The differences in $P_\ell$ due to differences in cloud-top 
    pressure $p_{\rm c}$ should be regarded with care:
    Fig.~\ref{fig:pressure_compare} shows the average polarization curves for
    $F_{\rm c}=0.5$ and different cloud-top pressures that were also shown 
    in Fig.~\ref{fig:pc_fc}, except here the shaded area represents the variability 
    of the 300~curves computed for $p_{\rm c}= 600$~mb.
    We have considered the variability within the $\pm 2\sigma$ interval, unless stated
    otherwise, where $\sigma$ is the (absolute) standard deviation of the distribution of
    values for $P_\ell$ obtained from the 300 generated cloud patterns, at a
    given $\alpha$.
    We used the $\pm 2\sigma$ interval in order to limit the influence of outliers.
    As can be seen in Fig.~\ref{fig:pressure_compare}, the variability due to 
    different cloud patterns for a given value of $F_{\rm c}$ is larger than the 
    differences in $P_\ell$ due to varying $p_{\rm c}$. 
    With patchy clouds, it is thus difficult to accurately retrieve $p_{\rm c}$
    from measurements of $P_\ell$ in a single wavelength region.
    Interestingly, the variability in the rainbow due to different cloud-top pressures appears to be negligible for every $F_{\rm c}$.

\begin{figure*}
        \centering
        \includegraphics[width=.6\linewidth]{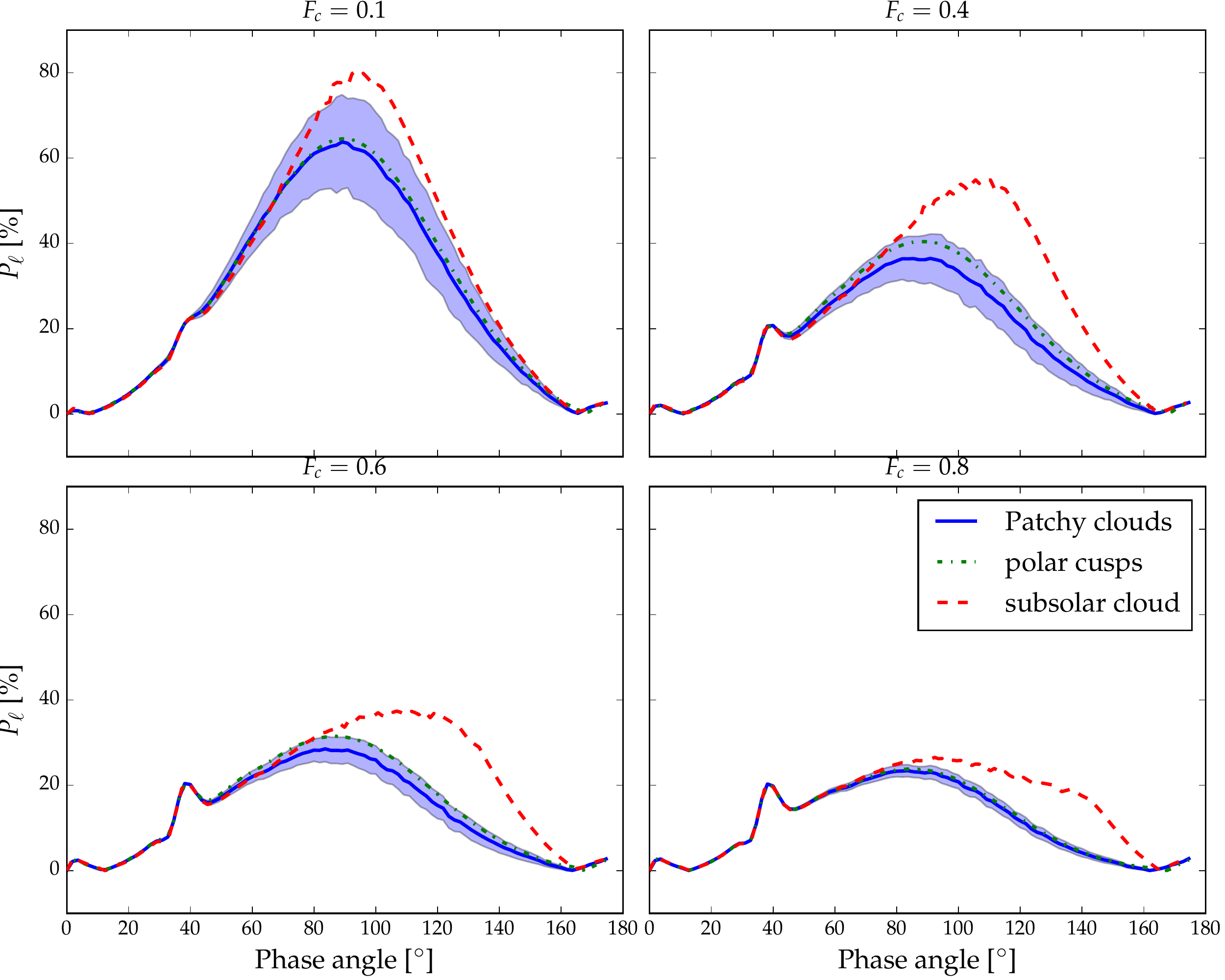}
        \caption{Comparison of $P_\ell$ at $\lambda=500$~nm for different
            types of cloud cover for $F_{\rm c}= 0.1$, 0.4, 0.6, and 0.8.
            The cloud-top pressure $p_{\rm c}$ is 800~mb.
            The solid line shows the average of 300 patchy cloud patterns. The shaded
            area shows the variability of the 300~curves.}
        \label{fig:compare_covers}
    \end{figure*}

\begin{figure*}
        \centering
            \includegraphics[width=.6\linewidth]{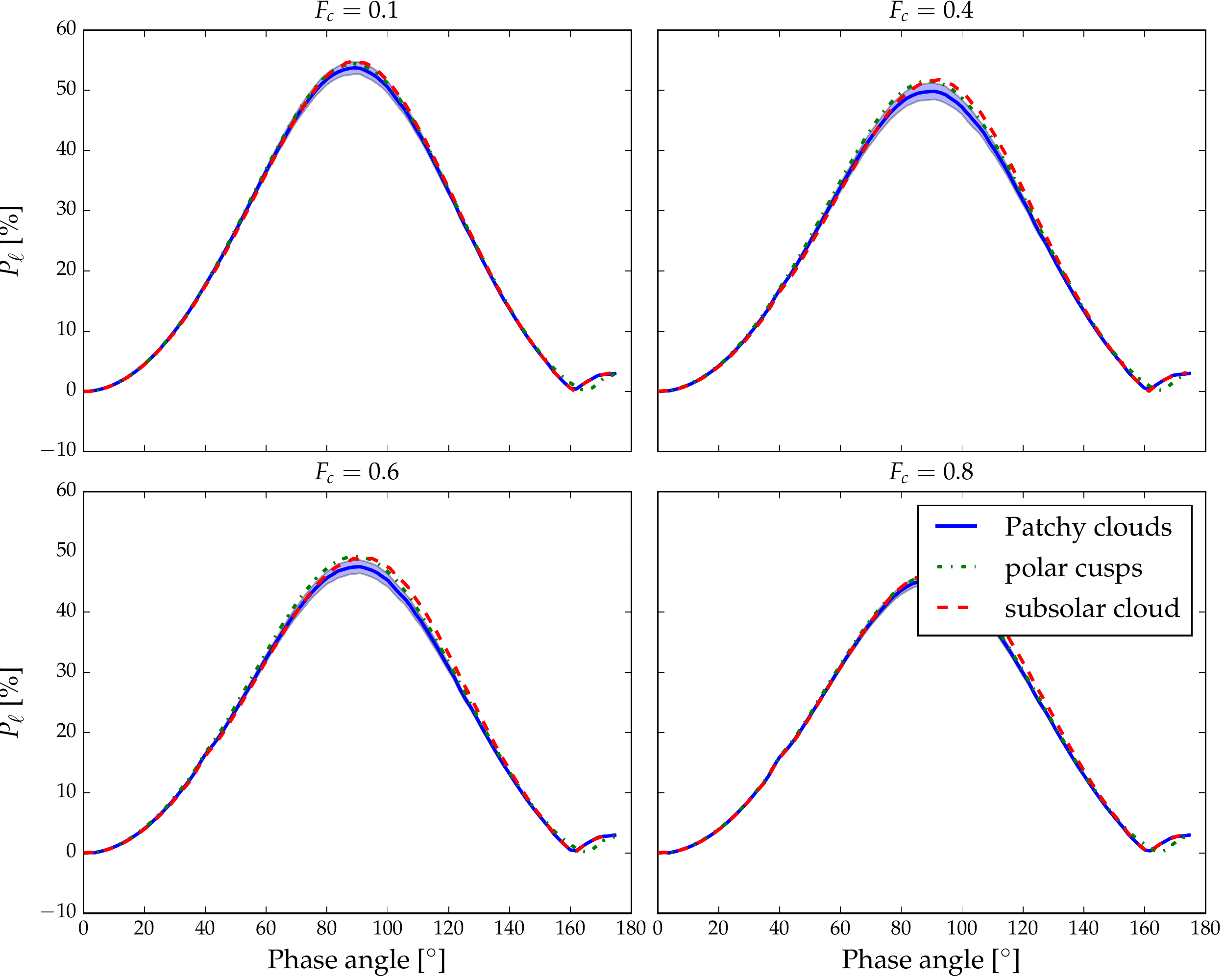} 
        \caption{Similar to Fig.~\ref{fig:compare_covers}, except at $\lambda=300~$nm.}
        \label{fig:blue_compare_covers}
    \end{figure*}

    The variability in $P_\ell$ depends on $F_{\rm c}$, 
    as shown in Fig.~\ref{fig:compare_covers}: the smaller
    $F_{\rm c}$, the larger the variability, because cloudy
    pixels have many more possible locations on the planet
    (the comparison between different cloud coverage types in this figure 
    will be discussed in Sect.~\ref{sec:comparison}). 
    With a small cloud coverage, the probability that the visible part of the
    planetary disk is completely cloud--free also increases, in particular 
    at larger phase angles. 
    Comparing Figs.~\ref{fig:compare_covers} and~\ref{fig:blue_compare_covers}, the
    effect of the wavelength $\lambda$ on the variability can be seen: at
    longer wavelengths (Fig.~\ref{fig:compare_covers}), the difference between
    the contribution of a cloudy and a cloud--free pixel is large. In the blue
    ($\lambda$=300~nm, Fig.~\ref{fig:blue_compare_covers}), however, the gas 
    above the clouds scatters more efficiently 
    and $P_\ell$ is less sensitive to the cloud distribution, resulting in less
    variability.

\begin{figure}
        \centering
        \includegraphics[width=0.9\linewidth]{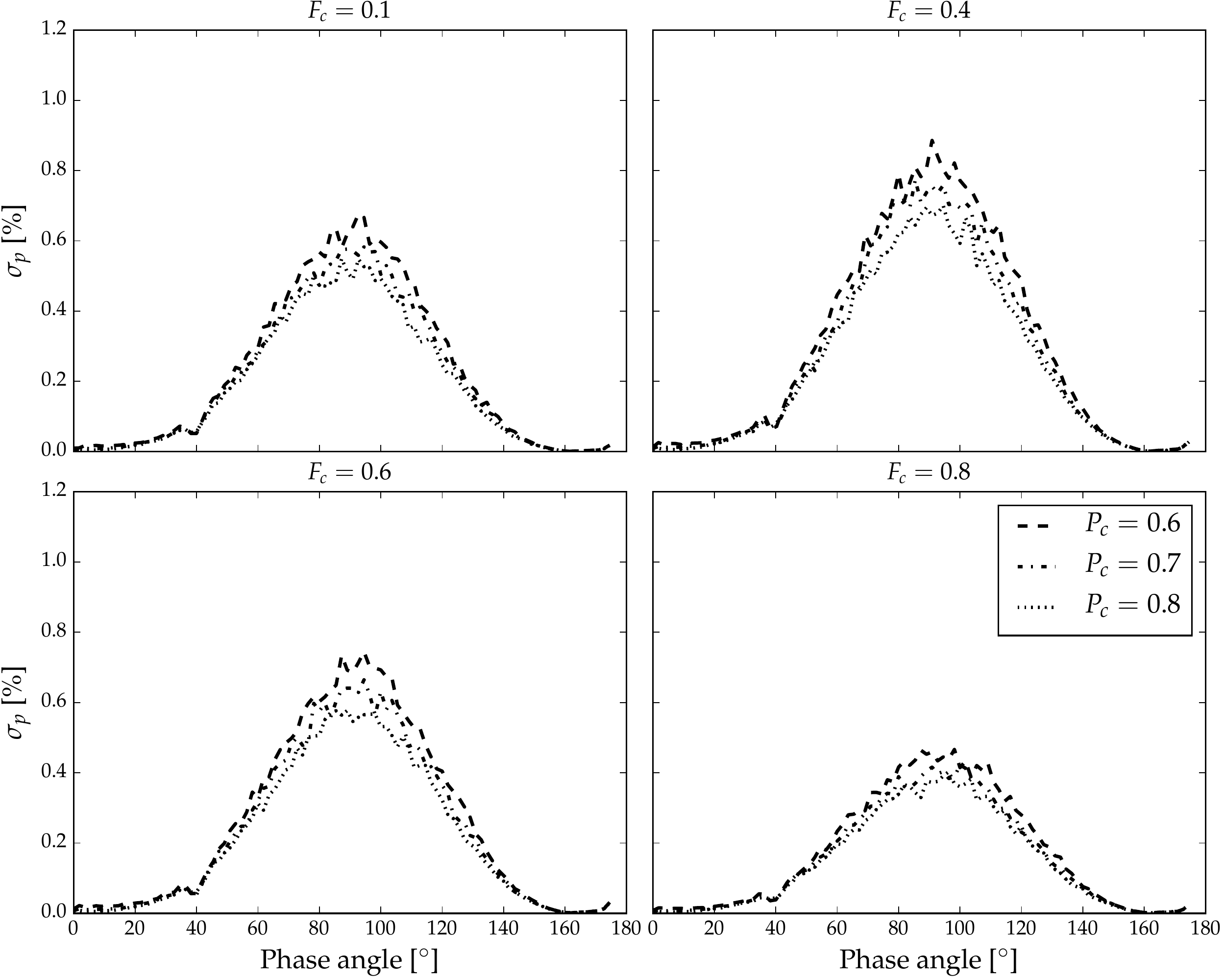}
        \caption{The 1-$\sigma$ variability on $P_\ell$ at $\lambda=300$~nm as a 
            function of $\alpha$ for different
            cloud coverages $F_{\rm c}$ and cloud-top pressures $p_{\rm c}$.}
        \label{fig:sigma-pressure}
    \end{figure}

    The dependence of the variability on the amount of Rayleigh scattering also 
    implies that with increasing cloud-top pressure $p_{\rm c}$ (i.e.,\ lower cloud-top altitude), the variability for a given value of $F_{\rm c}$ decreases.
    This shows from the $1\sigma$ variability of $P_\ell$ of planets with 
    patchy clouds as a function of $p_{\rm c}$ 
    (Fig.~\ref{fig:sigma-pressure}) in the blue ($\lambda=300$~nm).
    The $1\sigma$ variability of the flux is insensitive to $p_{\rm c}$.
    The variability in $P_\ell$ is not a direct proxy for cloud-top pressures and
    hence altitudes, because the measured variability will also be determined by 
    instrumental effects and observational constraints. Also,
    cloud patterns might not be as randomly located on a planetary disk 
    as in our model, and cloud-top pressures will vary across the planet.
    Nevertheless, our results imply that the variability of the polarization 
    could be a source of information.

    \begin{figure}
        \centering
        \begin{tabular}{c}
            \includegraphics[width=.75\linewidth]{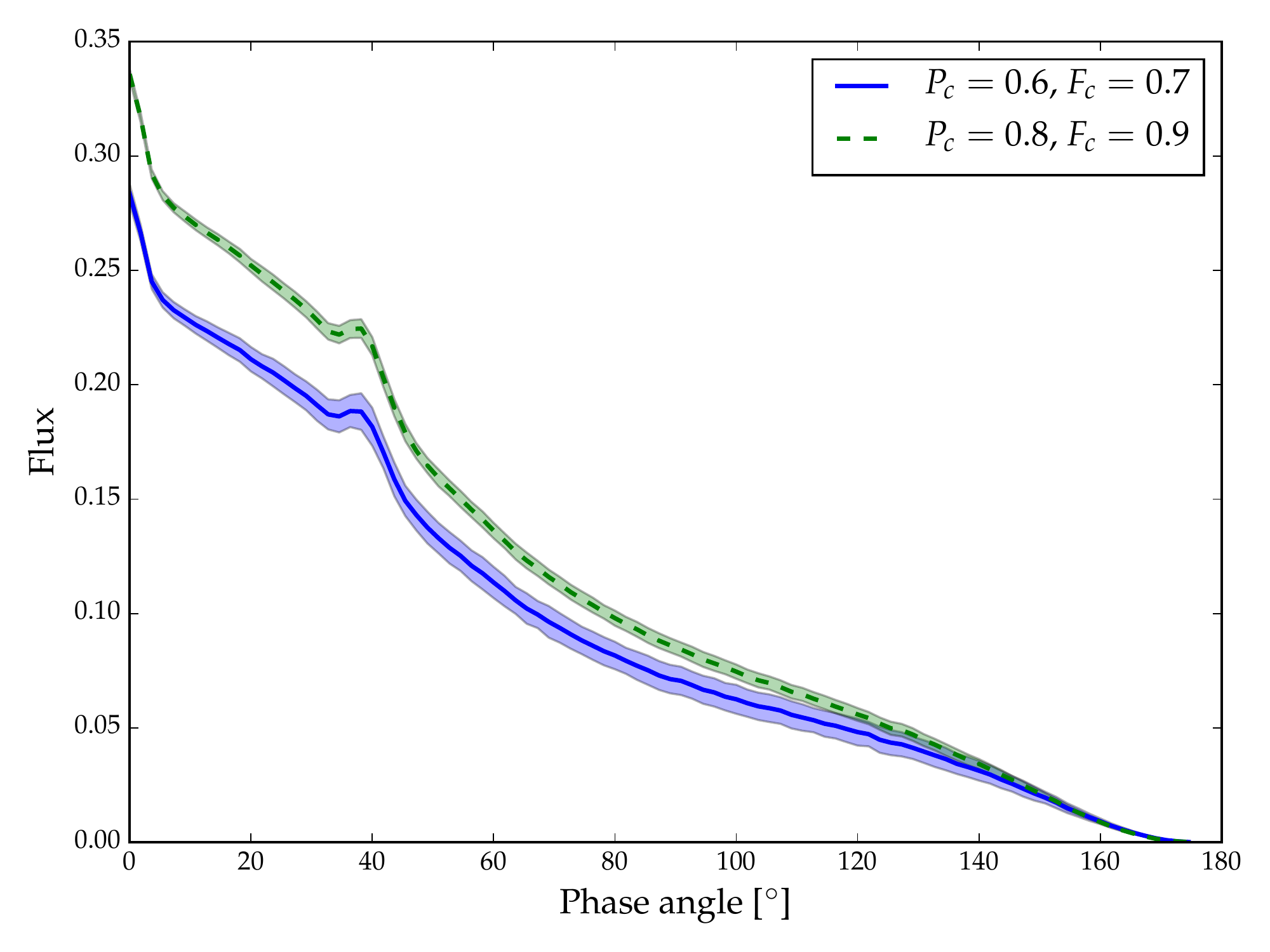}
            \\
            \includegraphics[width=.75\linewidth]{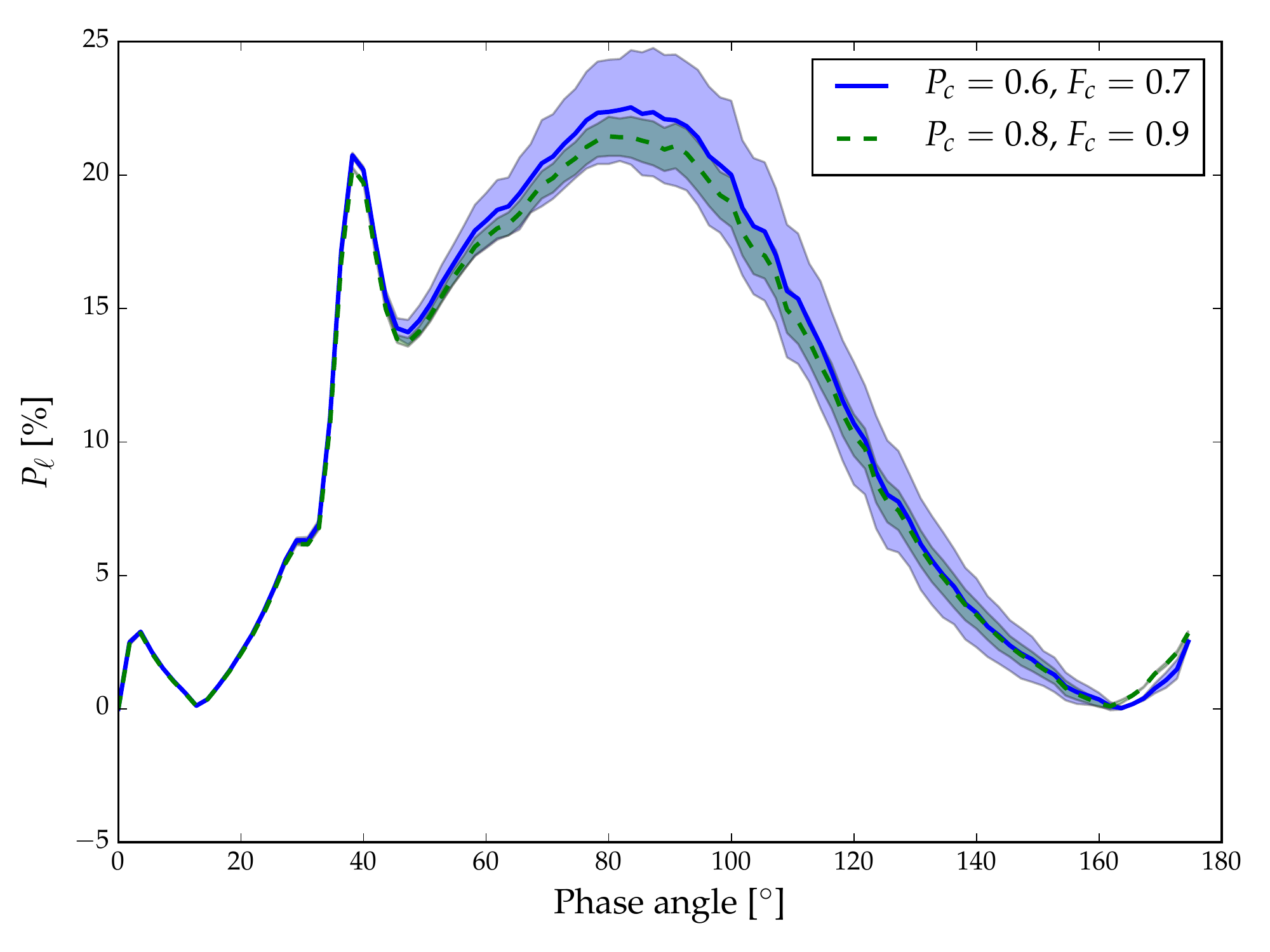}
        \end{tabular}
        \caption{The flux (top) and $P_\ell$ (bottom) at $\lambda=500~$nm 
            from two different patchy
            cloud configurations with $p_{\rm c}= 600~$mb, $F_{\rm c}=0.7$ (solid line)
            and $p_{\rm c}= 800~$mb, $F_{\rm c}=0.9$ (dashed line). The shaded areas
            represent the $2\sigma$ variability for 300 cloud patterns.}
        \label{fig:ambiguous_covers}
    \end{figure}

    Clearly, cloud-top pressure and cloud fraction both influence $P_\ell$
    and this could lead to ambiguous retrievals.
    As an example, Fig.~\ref{fig:ambiguous_covers} shows $P_\ell$ and its
    variability
    in the visible ($\lambda=500$~nm) for patchy clouds with $F_{\rm c}=0.7$ and
    $p_{\rm c}=0.6$~bar (case A), and for a cover with $F_{\rm c}=0.9$ and 
    $p_{\rm c}=0.8$~bar (case B).
    The variability of case B is completely contained in that of case A.
    This can be understood as follows: although case B has more clouds than
    case A, they are at lower altitudes, thus with a larger amount of
    Rayleigh scattering above the clouds, and mimicking the
    polarization of model A, but with less variability.

    If we now look at the total fluxes (Fig.~\ref{fig:ambiguous_covers}) for
    these two cases, the differences are much larger: as the cloud cover of case B
    is larger, this planet reflects more light than the planet of case A.
    More importantly, the variabilities of the two cases are mutually
    exclusive, especially for $\alpha< 90^\circ$. 
    Therefore, although the cases have similar polarization signals,
    they could be distinguished using their reflected flux, assuming the radius
    of the planet and/or its distances to its star and the observer are known 
    accurately enough. For example, to distinguish case A from case B using
    the planet's reflected flux, the planet radius 
    should be known well within 10~\%, assuming the distances are accurately
    known and the albedo of the surface below the clouds can be assumed to be
    similar (see Sect.~\ref{sec:surface}).

    We note that the strength of the rainbow is, again, nearly identical for both cases,
    both in flux (compared to the flux at slightly smaller or larger $\alpha$)
    and in polarization. This strengthens the application of the rainbow
    feature for cloud particle characterization
    \citep{Karalidi2012,Bailey2007}.


    \subsection{Comparing different cloud covers}
    \label{sec:comparison}

    We now compare the signatures of different types of cloud cover for the 
    same values of $F_{\rm c}$. 
    From Fig.~\ref{fig:compare_covers} ($\lambda=500$~nm),
    it seems that sub--solar clouds should be easiest to identify from
    measuring $P_\ell$ across a range of phase angles, 
    because $P_\ell$ will follow the 
    Rayleigh polarization curve once the cloud has disappeared over the limb. 
    The Rayleigh polarization curve is very distinct from the cloud polarization
    curve, at least for liquid water clouds.
    For example, in Fig.~\ref{fig:compare_covers}, the cloud coverage for
    $F_{\rm c}=0.10$ corresponds to a sub--solar cloud with $\sigma_{\rm c}= 20^\circ$
    (see Table~\ref{tab:eq-cloud-covers}), which disappears completely
    from view around $\alpha=100^\circ$.
    The cloud coverage for $F_{\rm c}=0.8$ corresponds to
    $\sigma_{\rm c}=60^\circ$ which disappears only around $\alpha=145^\circ$.

    The polar and patchy cloud types are more difficult to tell apart as they have a
    similar phase angle behavior. Polar clouds would, however, provide a more 
    stable signal than patchy clouds, especially for small values of $F_{\rm c}$, 
    and one could presumably use the variability
    as a proxy to determine the cloud type and to estimate the cloud cover
    patchiness.

    It is also worth noting that the three types of cloud cover all show a
    distinct rainbow in a phase angle region where the variability in $P_{\rm c}$ 
    due to cloud patchiness is small.
    This is important as the rainbow has been proposed as a tool to identify
    liquid water clouds, and to characterize their micro-physical properties
    \citep{Karalidi2011,Bailey2007}. 
    The rainbow can apparently not be used to retrieve 
    the type of cloud coverage, although the difference between $P_\ell$
    in the rainbow and in the continuum increases with increasing $F_{\rm c}$,
    for small ($< 0.6$) values of $F_{\rm c}$.
    
    Figure~\ref{fig:blue_compare_covers} is similar to Fig.~\ref{fig:compare_covers},
    except for $\lambda=300$~nm. At such short wavelengths, the
    scattering by the gas above the clouds obliterates any differences between
    the cloud coverage types. The maximum value of $P_\ell$ (around
    $\alpha=90^\circ$) decreases somewhat with increasing $F_{\rm c}$ 
    for all cloud coverage types. 
    The difference between the maximum $P_{\ell}$ obtainable for Rayleigh 
    scattering, and the maximum observed could thus help estimating $F_{\rm c}$.


        \subsection{Effect of the surface reflection}
    \label{sec:surface}
    
    In the simulations discussed so far, we have only considered black surfaces, 
    thus ignoring any reflection and (de)polarization that could be induced by light
    reflected off the surface. 
    On rocky planets, however, the surface could be covered by various types of rocks, sand, 
    liquids, soil, and even vegetation, thus with different surface albedos and/or 
    bidirectional reflection functions. 
    The influence of the albedo of Lambertian, thus completely depolarizing, reflecting surfaces
    on planetary phase curves has been investigated before 
    \citep[see, e.g.,][and references therein]{Stam2008}.   
    The main effect of such a reflecting surface is that it adds unpolarized light to 
    (the bottom of) the atmosphere, and thus usually increases the flux and decreases
    the degree of polarization of the light that emerges from the top of the atmosphere.
    The unpolarized surface light might change the angular location of the maximum 
    of polarization due to Rayleigh scattering, but not change the general phase angle 
    variation \citep{Stam2008}.
    As noted by \citet{Karalidi2012}, unpolarized surface reflection  might also 
    decrease the strength of the rainbow feature. 
    But in all cases, increasing the cloud coverage reduces the effect of 
    the surface on the phase curves in flux and polarization.
    
    The case of an ocean surface is less straightforward, because Fresnel reflection 
    is both anisotropic and polarizing. In particular, Fresnel reflection produces 
    the so--called sun-glint: the sharp reflection when the reflection angle equals the
    incident angle (and $\phi-\phi_0=0^\circ$). To investigate the effect of Fresnel reflection,
    we have performed similar computation as in the previous sections, except with
    cloud--free pixels in which the black surface is replaced by a Fresnel reflecting
    surface above a black water body, as also used by \citet{Stam2008}.
    We consider a calm, flat ocean to obtain the largest effect of the glint as waves 
    randomize and thus reduce the maximum of polarization due to Fresnel reflection 
    \citep{Zugger2010}.
    
\begin{figure*}
    \includegraphics[width=0.5\linewidth]{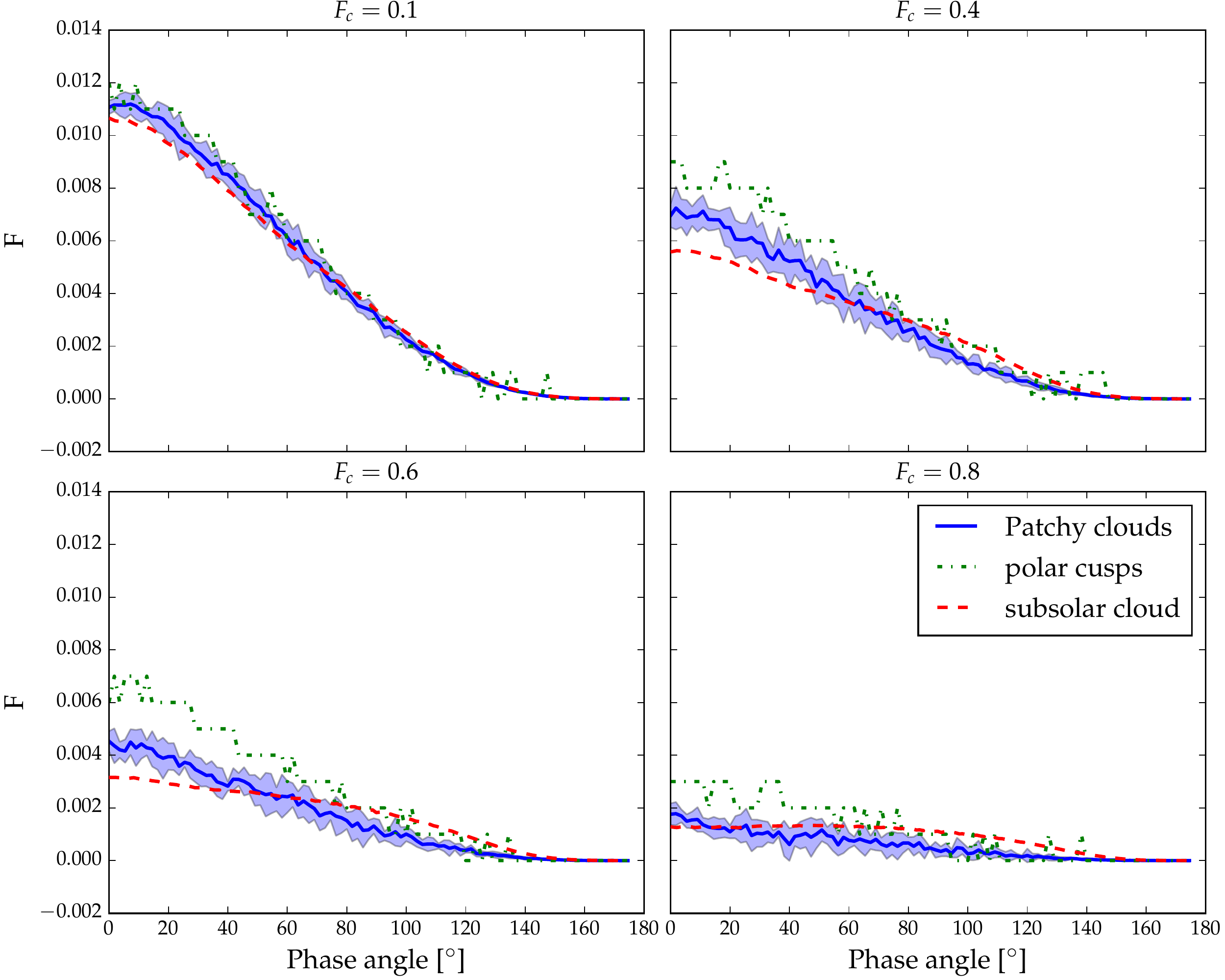}
    \includegraphics[width=0.5\linewidth]{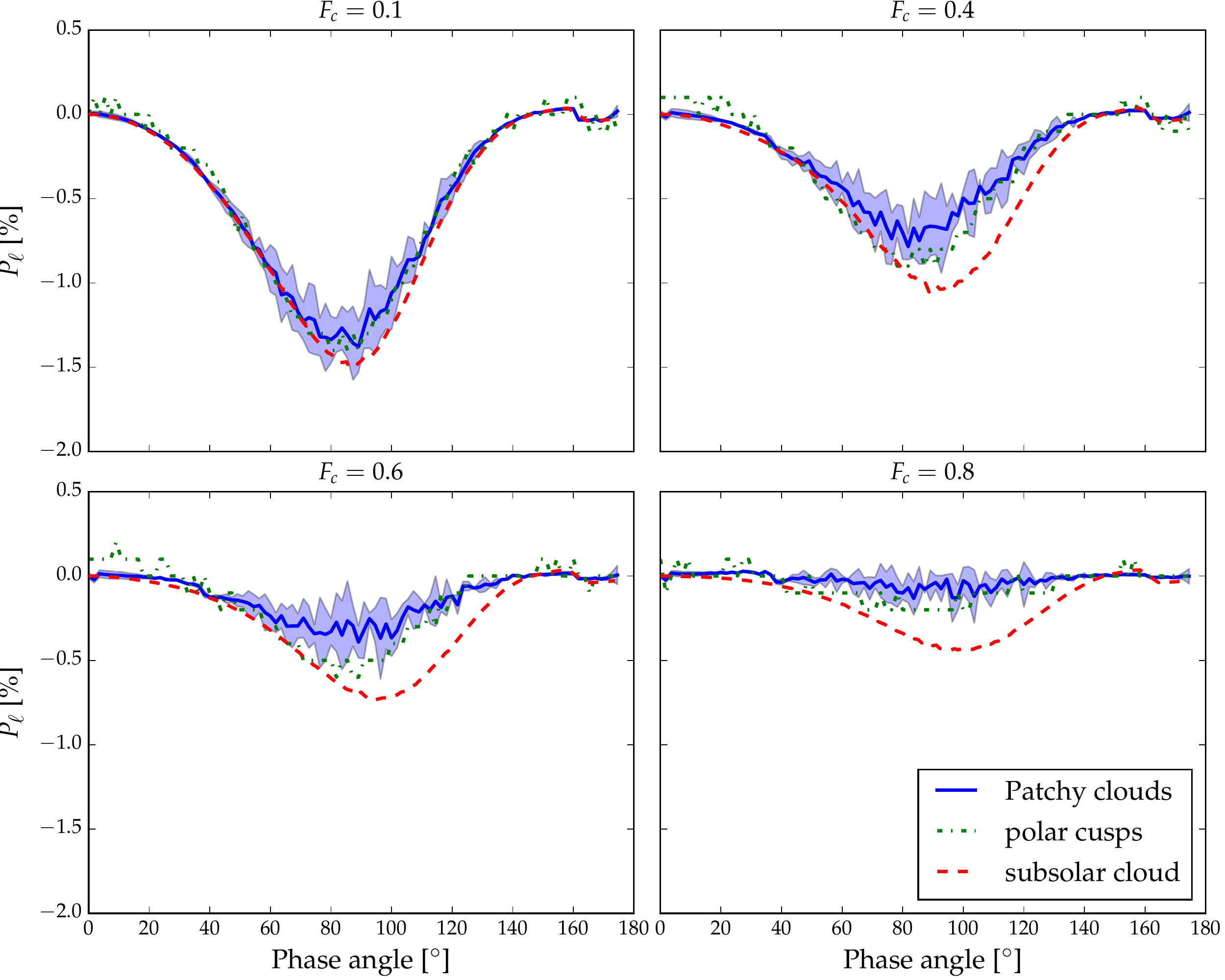}
    \caption{Differences $F^{\rm glint} - F^{\rm black}$ (left) and 
             $P^{\rm glint}_\ell - P^{\rm black}_\ell$
             (right) at $\lambda=300$~nm, as functions of the phase angle $\alpha$
             for different cloud coverage types and values of $F_{\rm c}$.
             }
    \label{fig:glint-I-300}
    \end{figure*}

\begin{figure*}
    \includegraphics[width=0.5\linewidth]{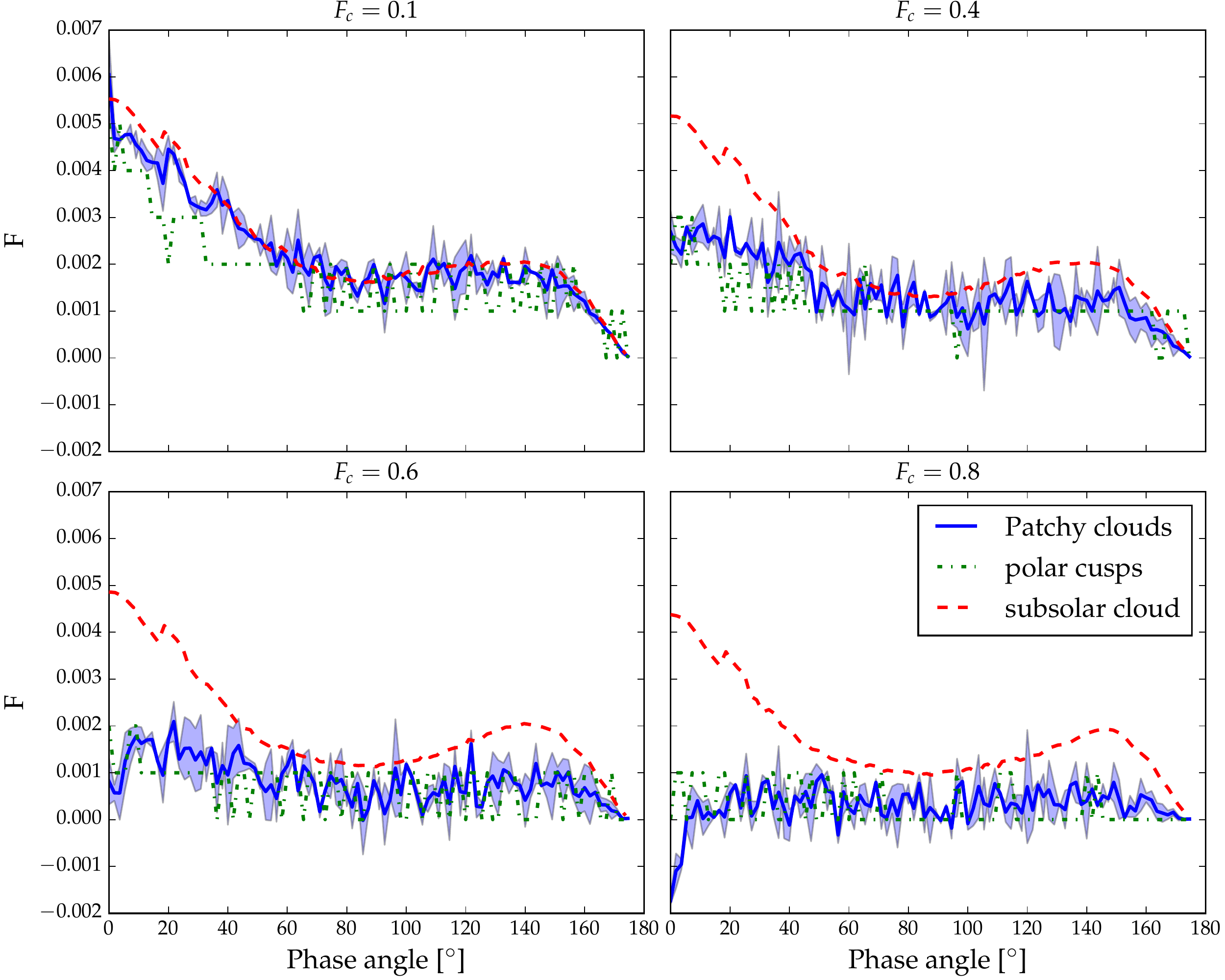}
    \includegraphics[width=0.5\linewidth]{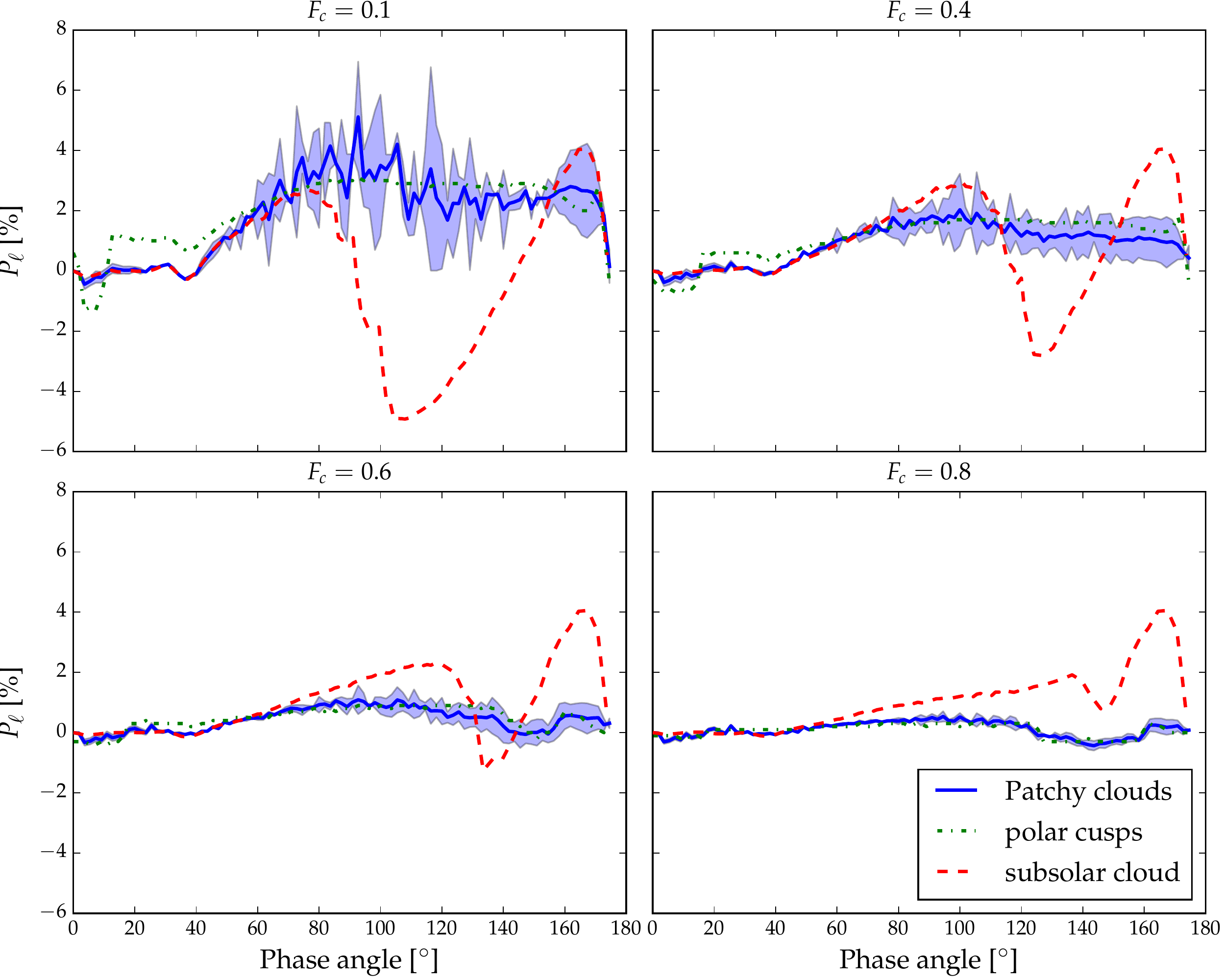}
    \caption{Similar to Fig.~\ref{fig:glint-I-300}, except for $\lambda=700$~nm.}
    \label{fig:glint-I-700}
    \end{figure*}

    The effects of the Fresnel reflection are very small, as can be seen in  
    Figs.~\ref{fig:glint-I-300}--\ref{fig:glint-I-700}, where we show the differences
    in the reflected flux and polarization at $\lambda=300$~nm and 700~nm. 
    The reflected flux is generally larger when 
    Fresnel reflection is included across the whole phase angle range.
    With increasing cloud coverage $F_{\rm c}$, the differences in flux 
    between the Fresnel reflecting surface and the black surface decrease, as expected.
    At $\lambda=300$~nm, the differences in flux are larger than at $\lambda=700$~nm;
    because the diffuse skylight is brighter at shorter wavelengths, and the surface
    receives and reflects light from and in all directions.
    Also, with increasing $F_{\rm c}$, the differences between the cloud
    coverage types increase, with in particular, the flux difference for the subsolar clouds
    being lower than for the other types at $\alpha=0^\circ$ and higher around 
    $\alpha=100^\circ$, but the differences are very small (about 1\% at 300~nm) 
    to start with. At 700~nm, the flux phase curve for the subsolar cloud case
    is clearly different from the other curves for $F_{\rm c} > 0.1$ 
    (although still the absolute differences are very small). 
    At large phase angles, the subsolar cloud disappears completely from sight 
    (even for $F_{\rm c}=0.8$), and the glint brightens the planet because the
    specular reflection of the direct starlight in the water tends to dominate 
    the disk--integrated signal when the planet is in a crescent phase \citep{Williams2008}.
    
    At 300~nm, the polarized signal of the planet with Fresnel reflection is lower than that of
    a black planet for almost all values of $\alpha$. Indeed, because at such short
    wavelengths, the surface is illuminated from all directions by the diffuse skylight
    and reflects back in all directions,
    the surface light decreases $P_\ell$ of the light emerging from the top of
    the atmosphere. The effect is largest for the subsolar cloud. 
    At longer wavelengths ($\lambda=700$~nm), where the Rayleigh scattering hardly 
    contributes, the effect of the Fresnel reflection is larger, and the Fresnel
    reflecting planet is more polarized than the black planet at almost all $\alpha$,
    except when there are subsolar clouds. The polarization differences for the 
    planets with subsolar clouds are actually
    very similar to those of the other planets up to the phase angle where the subsolar
    clouds disappear across the limb. While at moderate phase angles, the polarization
    of the planet with Fresnel reflection is still influenced by the Rayleigh scattering
    (see Fig.~\ref{fig:glint-I-300}), at larger phase angles, the polarization of 
    the glint (that is perpendicular to the reference plane) 
    becomes significant.
    The variability of patchy covers is barely affected by the presence 
    of the glint, with changes in values of $\sigma$ on a similar scale 
    as those due to the cloud-top pressure.
 
    \section{Observational strategy}
    \label{sec:obs-strat}
    Below, we propose an observational strategy that could allow us to 
    retrieve information about the cloud cover and the cloud properties. \\

    {\em Firstly}, in the blue ($\lambda < 400$~nm), the
    polarization is quite insensitive to the clouds because of the efficiency
    of Rayleigh scattering. 
    This is true for all types of cloud coverage (Fig.~\ref{fig:blue_compare_covers}). 
    Short wavelengths could thus be used to retrieve information on a planet
    that would be otherwise remain entangled with cloud properties and cloud coverage
    variations. 

    An example of such information is an estimation of a planet's orbital
    parameters, as this requires assumptions on the planet's polarization     
    \citep{Fluri2010},
    which in the visible is strongly influenced by for example, clouds.
    Changes in Stokes Q and U as a function of orbital phase, usually under the form of so--called $QU$--diagrams, that show linearly polarized flux $Q$ as a function
    of linearly polarized flux $U$ along the planetary orbit, 
    have been shown \citep{Brown1978, Wiktorowicz2015book}
    to be very helpful for the estimation of orbital parameters.
    Because clouds are still detectable in the blue as they do
    influence the maximum value of $P_\ell$ compared to that of 
    a cloud--free planet (see Fig.~\ref{fig:blue_compare_covers}),
    we have computed $Q$ and $U$ as a function of the planetary phase angle for some of the patchy cloud cases,
    for different values of the orbital inclination angle $i$ and the 
    longitude of the ascending node $\Omega$. 
    Angle $\Omega$ is defined as the angle between the observer's upward direction
    and the line through the two points of greatest elongation of a planet
    (for the description of the computation of the $QU$--diagrams, 
    see Appendix~\ref{app:rot}).

    \begin{figure}[h]
        \centering
        \includegraphics[width=0.9\linewidth]{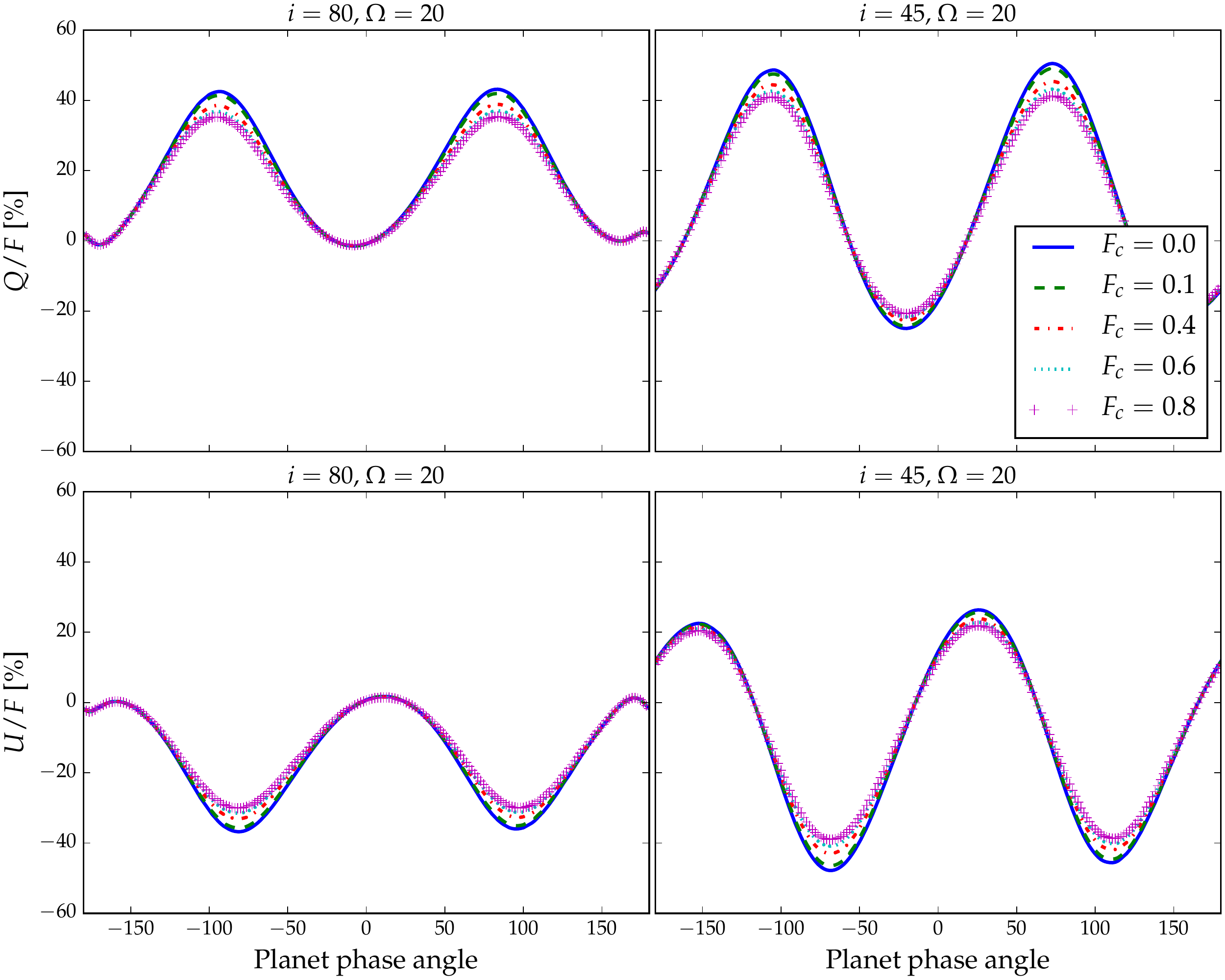}
        \caption{$Q/F$ (upper half) and $U/F$ (lower half) as a function of the planetary 
                phase angle at $\lambda=300$~nm
                for planets in circular orbits with patchy clouds and
            different cloud fractions $F_{\rm c}$. }
        \label{fig:QU-blue}
    \end{figure}

\begin{figure}[h]
        \centering
        \includegraphics[width=0.9\linewidth]{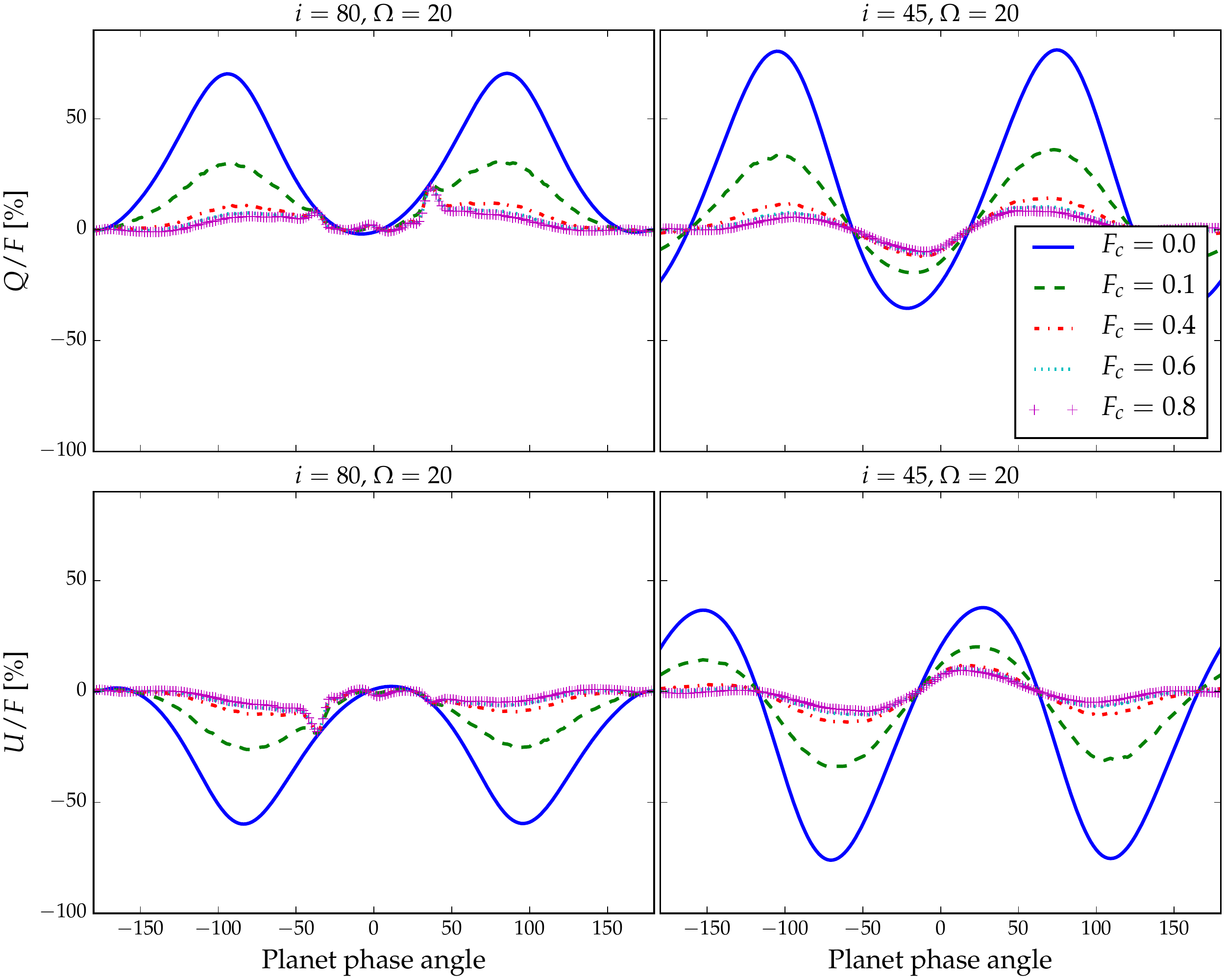}
        \caption{Similar to Fig.~\ref{fig:QU-blue}, except at 700~nm.}
        \label{fig:QU-green}
    \end{figure}

    Figure \ref{fig:QU-blue} shows $Q/F$ and $U/F$ as a function of planetary phase angle 
    (i.e., the phase angle of the planet when seen in an edge-on configuration, 
    see Appendix~\ref{app:rot} for its relation with the phase angle at other inclinations) 
    at 300~nm for two different
    values of $i$ and for different cloud fractions $F_{\rm c}$.
    At this short wavelength, the amplitude and orientation of the curves
    are insensitive to $F_{\rm c}$, indicating 
    the use of measuring (part of) this pattern for orbit parameter determination,
    without knowledge on the presence and/or distribution of clouds.
    Figure~\ref{fig:QU-green} is similar to Fig.~\ref{fig:QU-blue}, except at
    700~nm, where Rayleigh scattering is less efficient.
    Here, the influence of light that has been scattered by cloud particles results
    in a strong dependence of the curves on $F_{\rm c}$, 
    hence preventing the use of the determination of the orbital parameters.
    
    There is a caveat here: high altitude cirrus clouds have not been
    considered in this study.
    Although cirrus clouds can reach quite high altitudes, they are often
    optically thin \citep{Dupont2010} and their average coverage on
    Earth is quite small \citep[less than 20\%, according to][]{Rossow1999}.
    So it seems reasonable to assume that the no-clouds approximation would be
    valid in most cases for an Earth-like atmosphere.

        Other methods such as transits \citep{Seager2003}, transit time variations (TTVs), 
    astrometry \citep{Chauvin2012}, analysis of reflectance phase curves \citep{Kane2011} 
    and radial velocities could also provide estimates of the orbital parameters.
    
    {\it Secondly}, at any wavelength (but particularly in the visible) one can
    quickly infer whether the cloud cover is stable or not by measuring the
    variability of $P_{\ell}$ with time. Because $P_\ell$ is a relative 
    measure, this would require less stability of an instrument than when
    measuring variability in the reflected flux.
    If the variations of $P_{\ell}$ at a given phase angle are consistently
    greater than the measurement precision, and not periodic, 
    one could assume that the planet is covered with patchy clouds
    (although cloud coverage could partly be related to the distribution of
    continents on a planet, and thus partly periodic).\\

    {\it Thirdly}, the detection of optical phenomena such as rainbows and glories
    would give information about the microphysical properties of the clouds
    \citep{Karalidi2011, Karalidi2012, Rossi2015, Bailey2007}, although the
    glory would be difficult to detect on an exoplanet, because it occurs 
    in backscattered light and thus requires phase angles less than 10$^\circ$,
    \citep[see][]{GarciaMunoz2014}.
    The rainbow feature will be present for various types of (water) cloud cover,
    and a large range of cloud coverage fractions $F_{\rm c}$ and cloud-top pressures $p_{\rm c}$. \\

    {\it Fourthly}, when dealing with patchy clouds, it would be possible to use
    the reflected flux of the planet to constrain the cloud coverage fraction 
    and hence to retrieve unambiguously the cloud-top altitude. 
    In order to do this, the planetary flux should be well--calibrated over time,
    and the size of the planet and the distances to the star and the observer should be
    accurately known.
    We note that the cloud-top pressure could also be retrieved by studying the
    variability of the cloud cover in $P_\ell$, which shows some dependence on 
    $p_{\rm c}$ in the blue (cf. Fig.~\ref{fig:sigma-pressure}).
    For Earth--remote sensing, methods such as measuring the 
    reflected flux and/or polarization inside and
    outside of a gaseous absorption band are routinely used to retrieve
    cloud--top altitudes, and such methods could also be applied to exoplanets.
    However, applying those methods requires knowledge on the vertical distribution 
    of the absorbing gas \citep[see][and references
    therein for the application of this method to Earth-like exoplanets]{Fauchez2017}. \\
    
    {\it Fifthly}, once the type of coverage and the
    micro-physical properties of the clouds have been found, further fits of
    observations to models of the polarization could allow for a determination
    of the cloud fraction $F_{\rm c}$ within $10$~\% precision. \\

    Once the micro-physical properties of the clouds and the cloud fraction are
    obtained, these values could be used to make a refined estimation of the orbital
    parameters. Such an iterative process would hopefully converge toward a best--fit
    solution for both orbital parameters and cloud properties. \\


    
    All computations have been made assuming 
    phase angles from 0$^\circ$ to 180$^\circ$ are accessible for observations. 
    This range, however, depends both on the orbital inclination angle
    and on the inner working angle (IWA) of the telescope with the instrument
    (assuming observations that spatially resolve the planet from its star).
    Indeed, for an orbit that is observed under an inclination angle
    $i$, the range of phase angles the exoplanet goes through 
    along its orbit is given by 
    $90^\circ-i \leq \alpha \leq 90^\circ+i$
    (along an orbit with, for example, $i=30^\circ$, the smallest phase angle that
    the planet attains is thus 60$^\circ$ and the largest 120$^\circ$).
    The IWA limits the actual phase angle range at which an exoplanet can
    be separated from its star, cutting off access to the 
    smallest and largest phase angles, where the planet is too close to its star
    to be resolved. We note that the IWA will usually depend on the wavelength.
    
\begin{figure}[h]
        \centering
        \includegraphics[width=\linewidth]{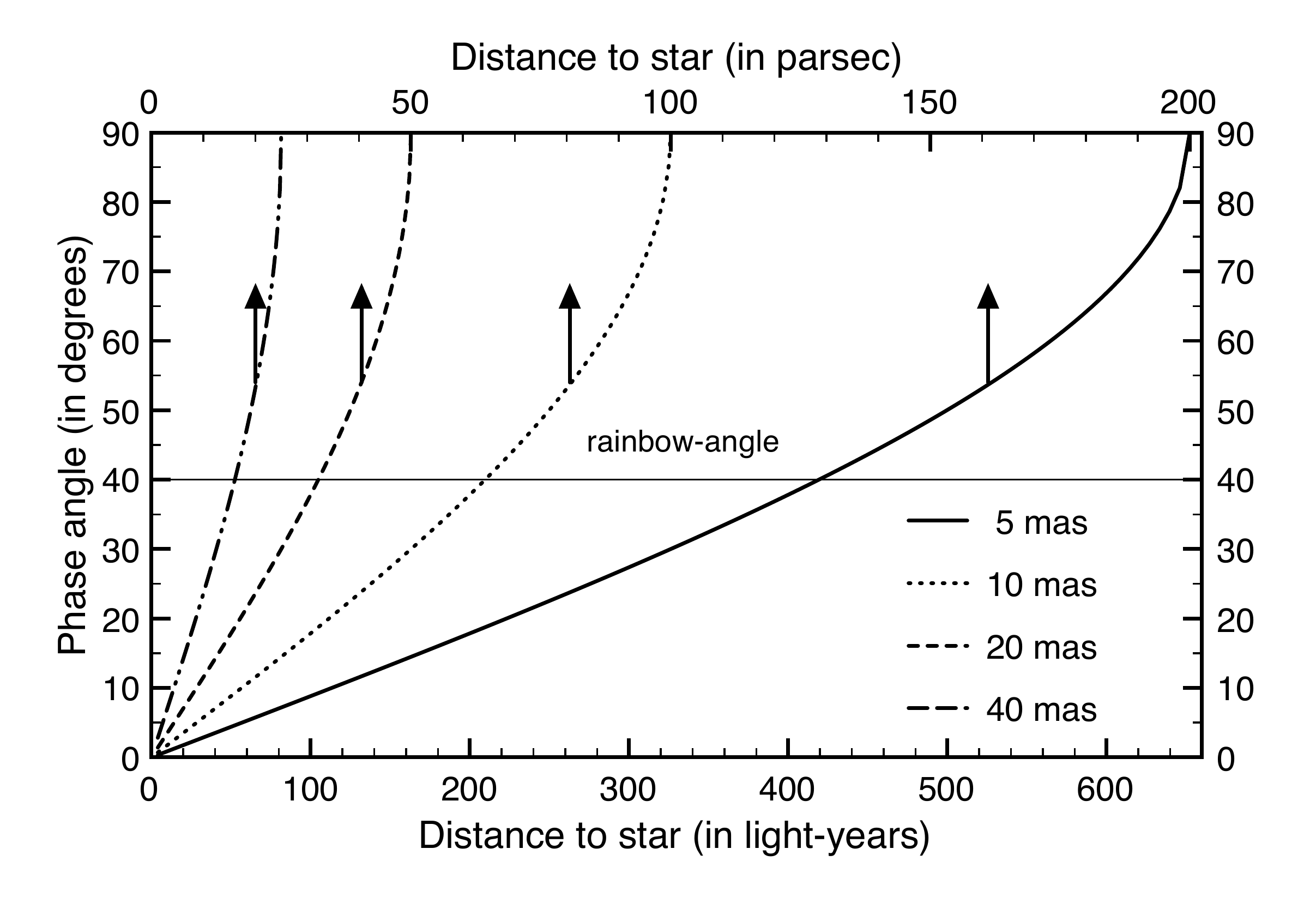}
        \caption{Observable phase angles for a planet at 1~AU of its star as a
        function of the observer's distance to the star. Different lines show
        results for different inner working angles (IWA). The solid line at 
        $\alpha=40^\circ$ corresponds to the angle at which the rainbow appears,
        assuming water droplets. For the phase angle range 90$^\circ$ - 180$^\circ$,
        the graph mirrors this one over $\alpha=90^\circ$.}
        \label{fig:iwa}
    \end{figure}

    In Fig.~\ref{fig:iwa}, we show the phase angle range at which a planet at a
    distance of 1~AU from its star (i.e., the habitable zone
    around a solar-type star) can be observed as a function of the distance between the 
    observer to the star for IWAs of 5, 10, 20, and 40 mas, ignoring 
    the limitations of a planet's orbital inclination angle $i$.
    For example, with an IWA of 40 mas, a planet at 1 AU from a star at 200 pc,
    can only be observed at phase angles larger than about 20$^\circ$.
    Whether or not the planet will actually present itself at this phase
    angle at some point in time, will depend on the planet's orbital inclination angle.
    With an IWA of 20 mas, a planet at 1 AU from its star cannot be spatially
    resolved if the system is at larger distances than about 160 light-years
    (50 parsecs). Also shown in the figure is the phase angle of 40$^\circ$ 
    around which the primary rainbow,
    indicative for light scattered in liquid water cloud droplets
    \citep[][]{Karalidi2012,Bailey2007} would be visible. 
    For example, with an IWA of 20 mas, the
    rainbow would be observable for systems that are closer than about 100
    light-years, of course provided $i > 50^\circ$. 
    For systems that are further out, the planet will too close to its star at
    the rainbow phase angle to be spatially resolved from its star with this IWA.

    \section{Conclusion}
    \label{sec:conclusion}

    Identifying and characterizing clouds on exoplanets is crucial for
    retrieving their atmospheric properties and for getting insight in their
    climate and habitability.  
    We have shown that polarimetry could enable observers to derive
    information on the type and fraction of cloud coverage. 
    In our modeling, we have concentrated on Earth--like exoplanets with black
    surfaces with and without a Fresnel reflecting interface. The latter would
    be representative for a flat ocean surface.
    While similar conclusions can be expected for gaseous planets, the effects
    of deeper cloud layers, with a possible strong vertical variation in cloud 
    particle micro-physical properties, remain to be studied.

    Polarimetry allows us to distinguish sub-solar clouds from patchy and
    polar clouds, because at a phase angle that depends on the cloud's spatial 
    extension, sub-solar clouds will disappear (and reappear) over the limb of 
    the planet, leaving (and removing) the characteristic
    polarization signature of Rayleigh scattering gas.
    Secondly, the variability of the polarization signature of patchy clouds
    should allow us to distinguish them from polar clouds, as the latter exhibit
    less variability at all phase angles.
    Measurements of the variability of the polarization combined with accurate
    measurements of the planet's reflected flux (which requires knowledge about the
    size of the planet and its distance to the parent star) would provide
    a tool to reduce ambiguities between the fraction of patchy clouds and the
    cloud-top pressure, as the polarization variability due to the varying
    patchiness appears to be larger than that due to the cloud-top pressure.
    
    Finally, measurements at short wavelengths ($< 400$~nm) would allow
    the observer to mostly ignore the effect of the clouds on the planet's 
    polarization signal 
    and would therefore allow us to characterize the gaseous atmosphere of the
    planet (down to the cloud tops).
    At these wavelengths, a pure Rayleigh scattering atmosphere approximation
    could also be used to derive orbital parameters without too much interference
    of the clouds.
    Longer wavelengths could be used to estimate the cloud coverage, and,
    depending on the phase angle and the IWA of the telescope + instrumentation, 
    to derive micro-physical properties from observations of the primary rainbow,
    provided the rainbow phase angle ($\alpha=40^\circ$) is within reach
    for the planetary system under study.


\begin{acknowledgements}
    LR thanks Emmanuel Marcq and Arnaud Beth for their useful comments
    regarding an earlier version of this work.
    We also thank the reviewer who greatly helped improving the paper.
    LR acknowledges the support of the Dutch Scientific Organization (NWO)
    through the PEPSci network of planetary and exoplanetary science.
\end{acknowledgements}

%
%

\bibliographystyle{aa} 
\bibliography{biblio.bib} 

\begin{appendix} 

    \section{Influence of computation parameters}
    \label{app:npix}

    \subsection{The number of pixels $n_{\rm pix}$}

    The number of pixels across the planet directly determines the spatial resolution
    on the planetary disk, and thus the size of the cloudy pixels on the disk. 
    A large number of pixels allows us to take smaller spatial features into account but 
    also leads to long computation times.

\begin{figure}[h]
        \centering
        \includegraphics[width=0.75\linewidth]{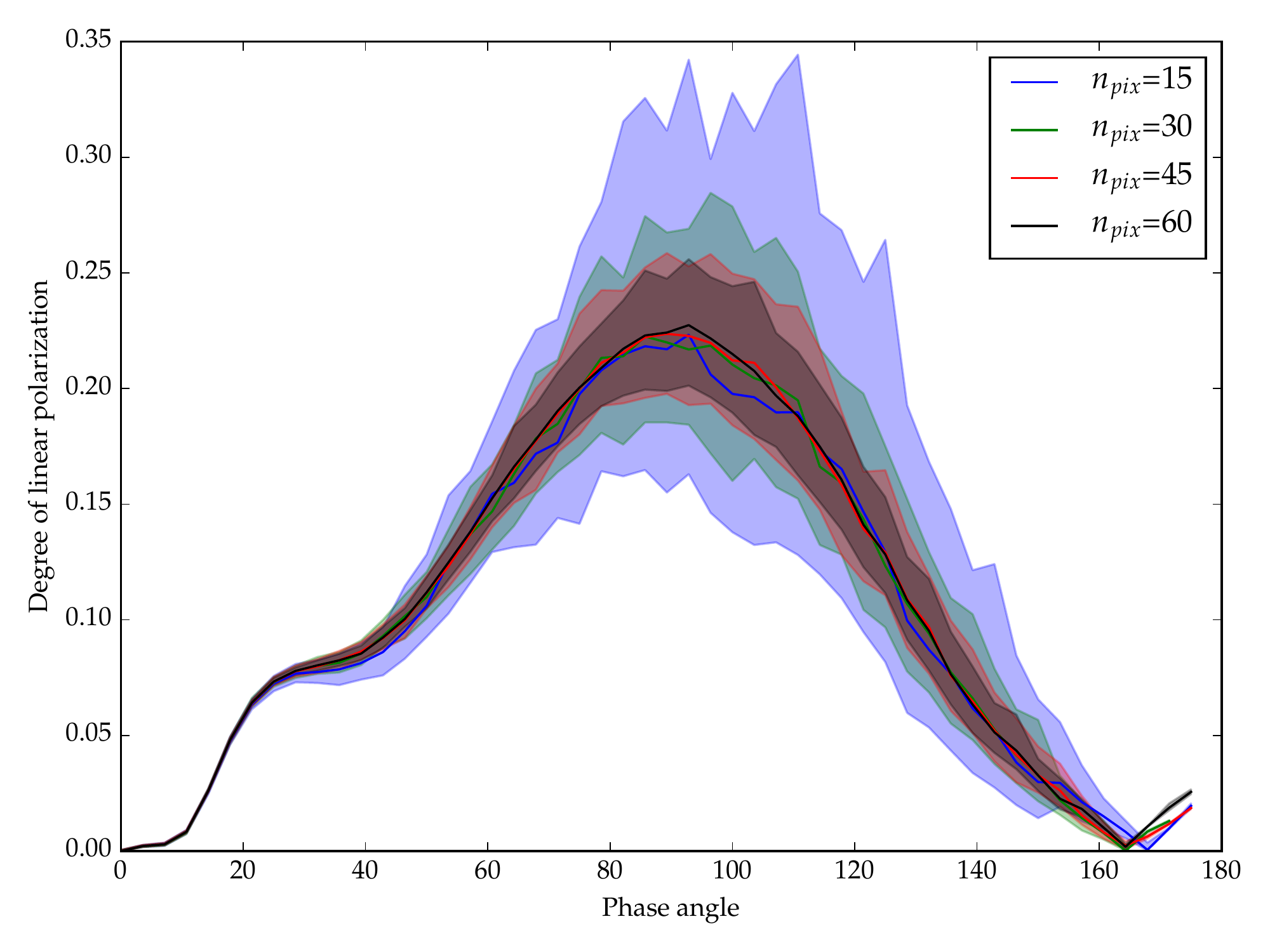}
        \caption{Degree of linear polarization $P_\ell$ as a function of the
        phase angle $\alpha$ for a $50\%$ patchy cloud coverage. The solid curves 
        represent the average of $P_\ell$ over 100~cloud patterns.
        The shaded areas correspond to the $2\sigma$ variability.}
        \label{fig:npix-effect}
    \end{figure}

    To measure the influence of $n_{\rm pix}$ on the computed degree of polarization
    $P_\ell$, we have run simulations with a $50$\% patchy cloud cover ($F_{\rm c}=0.50$) 
    for 100~cloud patterns and different values of $n_{\rm pix}$. 
    The result is shown in Fig.~\ref{fig:npix-effect}, in which we indicate the $2\sigma$ variability of $P_\ell$.

    It can be seen that the main effect of increasing $n_{\rm pix}$ is the reduction
    of the variability, especially when increasing from $n_{\rm pix}=15$ to $n_{\rm pix}=30$.
    Overall, using a smaller value for $n_{\rm pix}$, and thus larger pixels
    on the planetary disk, leads to more abrupt pixel type differences
    across the disk and therefore to more variability in the polarization.
    The decrease of the variability with increasing value of $n_{\rm pix}$,
    and thus smaller pixels, seems to have converged with $n_{\rm pix}=45$ in the figure.
    As a compromise between a good enough accuracy and a reasonable computation
    time, we decided to pursue the calculations for this paper with $n_{\rm pix}=40$.


    \subsection{The number of cloud patterns $n_{\rm pattern}$}

\begin{figure}[h]
        \centering
        \includegraphics[width=0.75\linewidth]{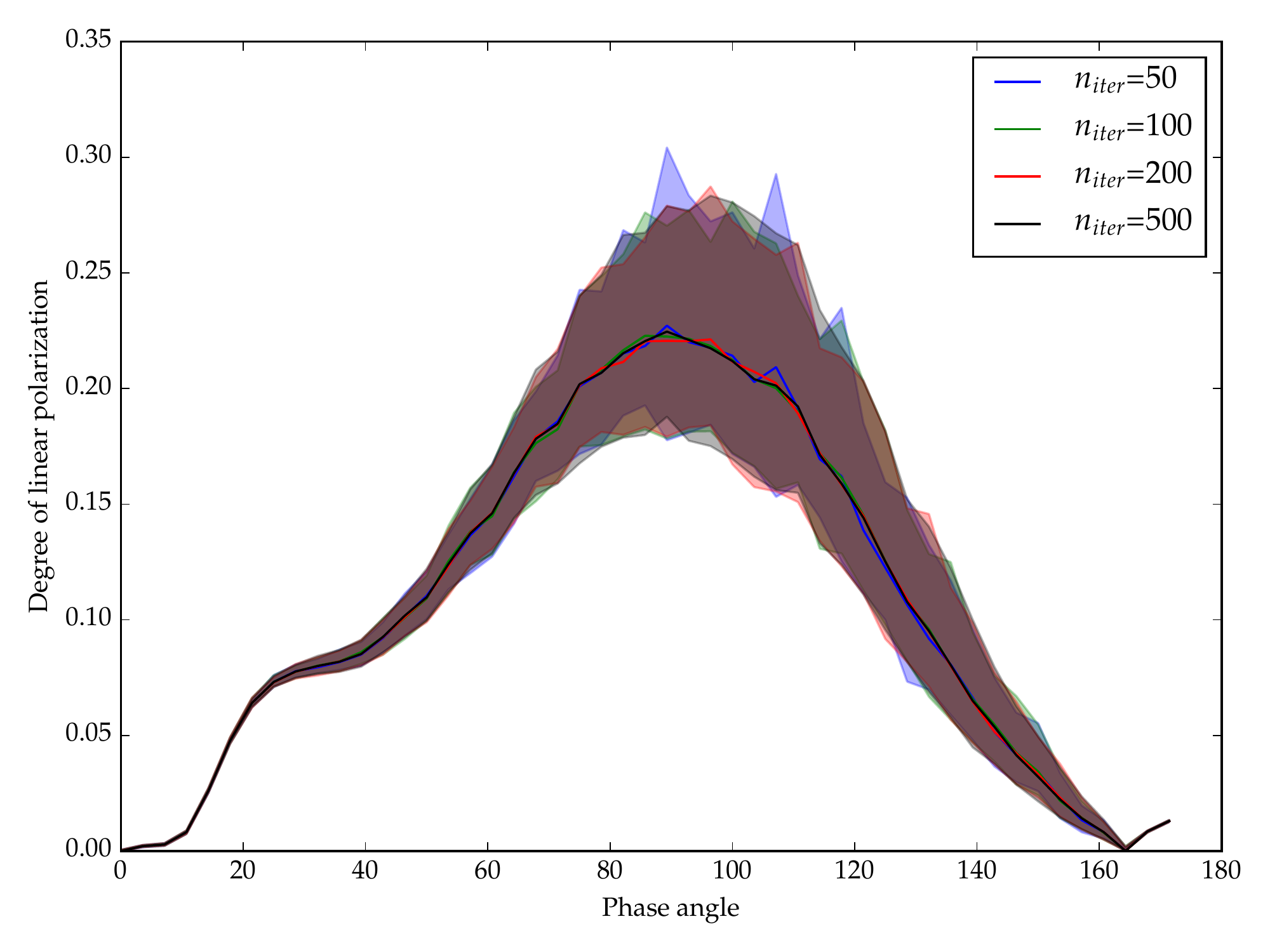}
        \caption{Similar to Fig.~\ref{fig:npix-effect}, except for $n_{\rm pix}=40$
        and different numbers of cloud patterns $n_{\rm pattern}$.}
        \label{fig:niter-effect}
    \end{figure}

    For patchy clouds and a given value of the cloud coverage $F_{\rm c}$,
    the number of cloud patterns, $n_{\rm pattern}$, might influence the computed average 
    value of $P_\ell$ and its variability, so it is 
    necessary to find the minimum number of patterns required for accurate results.
    Figure~\ref{fig:niter-effect} shows $P_\ell$ for $F_{\rm c}=0.50$ 
    and $n_{\rm pix} = 40$ for different values of $n_{\rm pattern}$.
    As can be seen in the figure, $n_{\rm pattern}$ has only a small influence
    on $P_\ell$ and its variability: increasing $n_{\rm pattern}$ smoothens 
    the average curve and the variability.
    We have performed our computations with $n_{\rm pattern}= 300$ to ensure
    representative results without being burdened by long computing times.

\end{appendix}

\begin{appendix}

\section{Rotating polarization reference planes}
\label{app:rot}

\subsection{From local to disk--integrated Stokes vectors}

We compute ${\bf F}_{\rm planet}$, the disk--integrated Stokes vector of a 
model planet, by summing up Stokes vectors computed for $N$ locations on 
the illuminated and visible planetary disk, as follows
\begin{equation}
   {\bf F}_{\rm planet} = \Sigma_{i=1}^{N} {\bf F}_i \hspace*{0.1cm} a_i,
\label{eqflux}
\end{equation}
where ${\bf F}_i$ is the locally reflected Stokes vector, and $a_i$ the surface area 
of pixel $i$ on the two dimensional planetary disk. The center of pixel $i$, 
projected parallel to the line of sight onto the three dimensional planet indicates
the location from where ${\bf F}_i$ has been reflected.

Assuming that the planet presents a circular disk to the observer with a 
radius equal to 1.0, the following should hold
\begin{equation}
   \Sigma_{i=1}^{N} a_i = \pi.
\end{equation}
We divided the planetary disk in areas that are equal in size. 
Equation~\ref{eqflux} thus transforms into
\begin{equation}
   {\bf F}_{\rm planet} = \pi \hspace*{0.1cm} \Sigma_{i=1}^{N} {\bf F}_i.
\label{eqfluxi}
\end{equation}
Before evaluating this summation, we had to make sure that all locally reflected
Stokes vectors are defined with respect to the same reference plane.

Our radiative transfer algorithm provides parameters $Q_i$ and $U_i$ of each 
locally reflected Stokes vector as defined with respect to the 
\emph{local meridian plane}: 
the plane through the local directions toward the zenith and the observer. 
We note that this local meridian plane is independent of the direction toward
the parent star.
The natural reference plane for the disk--integrated Stokes vector ${\bf F}_{\rm planet}$
is the \emph{planetary scattering plane}: the plane through the center of the planet,
the sun and the observer. The advantage of the planetary scattering plane is 
that when the planet is mirror--symmetric with respect to this reference plane,
disk--integrated Stokes parameter $U_{\rm planet}$ will equal zero (disk--integrated
circular polarization Stokes parameter $V_{\rm planet}$ will then also equal zero).

To rotate from one reference plane to another, we used rotation matrix ${\bf L}$,
which is given by \citep[see][]{Hovenier1983}
\begin{equation}
   {\bf L}(\beta)= \left[ \begin{array}{cccc}
             1 & 0 & 0 & 0 \\
             0 & \cos 2\beta & \sin 2\beta & 0 \\
             0 & -\sin 2\beta & \cos 2\beta & 0 \\
             0 & 0 & 0 & 1 \\
\end{array}
\right],
\label{eq_L}
\end{equation}
with $\beta$ the angle between the two reference planes, measured rotating in the
clockwise direction from the old to the new reference plane when looking
toward the planet ($0^\circ \leq \beta < 180^\circ$) (\citet{Hovenier1983}
write that $\beta$ is measured rotating in the anti--clockwise direction 
when looking toward the observer, which of course yields the same angle).

\begin{figure}[h]
   \centering
   \includegraphics[width=1.0\linewidth]{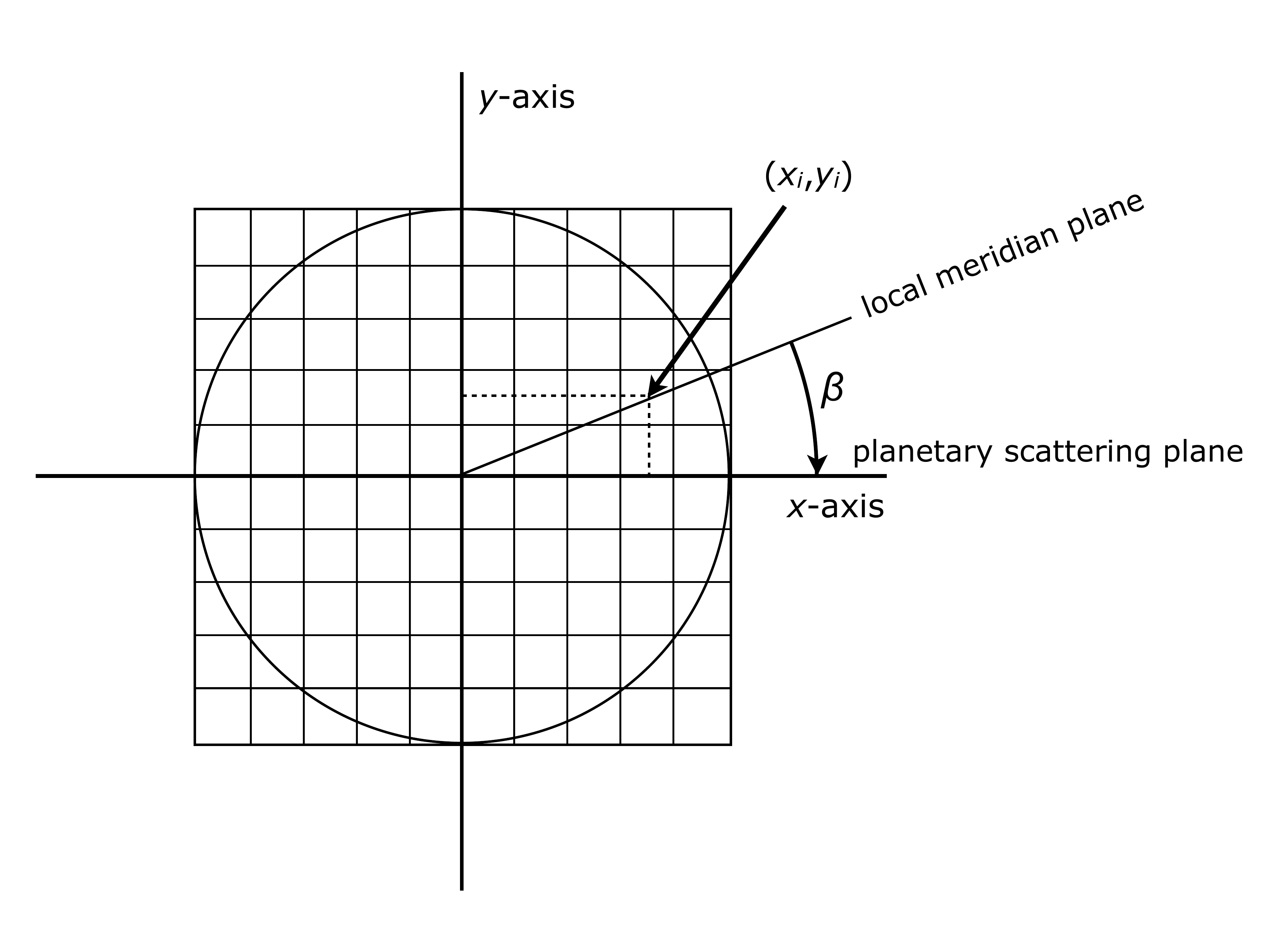}
   \caption{The definition of rotation angle $\beta$ for a pixel with 
            center coordinates $(x_i,y_i$) on
            the planetary disk.}
\label{fig:b1}
\end{figure}

Rotation angle $\beta_i$ for a given pixel $i$ depends on its
location with respect to the planetary scattering plane. In the following, we 
use a Cartesian $xy$-coordinate system, with the origin in the center
of the disk and the $x$--axis horizontal through the disk (see Fig.~\ref{fig:b1}). 
The radius of the disk equals 1.
We can distinguish the following cases given pixel center 
coordinates $(x_i,y_i)$: \\

\noindent $x_i \hspace*{0.1cm} y_i \geq 0$ :
$\beta_i = \arctan{ \left( y_i/x_i \right)}$ \\

\noindent $x_i \hspace*{0.1cm} y_i < 0$ :
$\beta_i = 180^\circ + \arctan{\left( y_i/x_i \right)} $ \\

\subsection{From planetary scattering plane to detector plane}

\begin{figure}[h]
   \centering
   \includegraphics[width=1.0\linewidth]{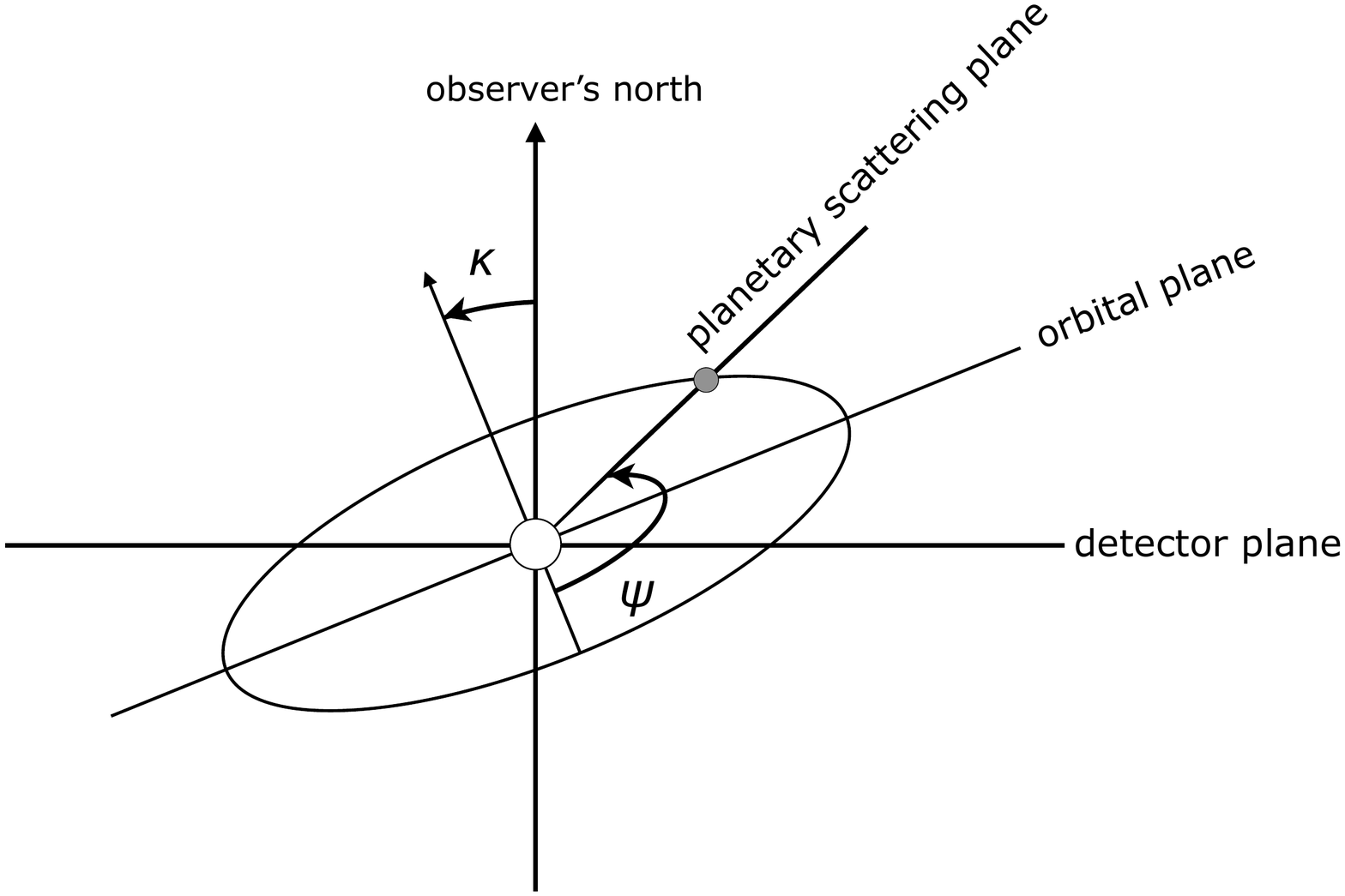}
   \caption{An inclined, circular planetary orbit with the planet's orbital position 
            angle indicated by angle $\psi$ ($0^\circ \leq \psi \leq 360^\circ$). 
            The rotation of the normal on the 
            planetary orbit as projected on the sky with respect to the observer's 
            north is indicated by angle $\kappa$ ($-90^\circ \leq \kappa \leq +90^\circ$).}
\label{fig:b2}
\end{figure}

The orientation of the planetary scattering plane with respect to the observer 
depends on the inclination angle $i$ of the planetary orbit, on the angle $\kappa$
between the observer's upward direction and the projection of the normal on the planetary 
orbital plane on the sky, and on the
position angle $\psi$ of the planet along its orbit (see Fig.~\ref{fig:b2}).
The longitude of the ascending node $\Omega$ equals $90^\circ$ - $\kappa$.
Below, we derive how the Stokes vector ${\bf F}_{\rm planet}$ that has
been computed for the planet as a whole and with respect to the planetary
scattering plane, can be rotated to an observer's reference plane.
For the latter we have used a horizontal plane, that we refer to as the
detector plane.

The orbital inclination angle $i$ is defined as the angle between the normal on the
planetary orbit and the direction toward the observer. 
The inclination angle has a value between $0^\circ$
(for a 'face--on' orbit) and 90$^\circ$ (for an 'edge--on' orbit). 
In the following, we use $\kappa=0^\circ$, assuming that the observer's 
telescope and detector are rotated to accomodate this (see Fig.~\ref{fig:b2}).
The normal on the planetary orbital plane thus falls in the
plane that is perpendicular to the detector plane and that contains 
the direction toward the observer.

The planet's  orbital position angle $\psi$ is measured from the position
where the planet is closest to the center of the stellar disk as seen by
the observer. Thus, for $i=90^\circ$, $\psi= 0^\circ$ in the middle of 
the primary transit, and $\psi=180^\circ$ in the middle of the secondary transit.
For $i=0^\circ$, $\psi=0^\circ$ is undefined.
Angle $\psi$ is measured rotating from $\psi=0^\circ$ in the counter--clockwise direction
(for a planet orbiting in the clock--wise direction as seen by the 
observer, $\psi$ will thus decrease in time).

For completeness, given the orbital inclination angle $i$ and the orbital position
angle $\psi$, the planetary phase angle is given by
\begin{equation}
   \alpha = \arccos \left(-{\cos{\psi} \sin{i}} \right).
\label{eq_alpha}
\end{equation}

Angle $\beta$ to rotate the planetary Stokes vector from the planetary
scattering plane to the detector plane depends on the orbital position 
angle $\psi$ and the orbital inclination angle $i$: \\

\noindent $\tan{\psi} \geq 0.0$ :
$\beta = 180^\circ - \arctan{\left( \cos{i} / \tan{\psi} \right)}$ \\

\noindent $\tan{\psi} < 0.0$ :
$\beta = - \arctan{\left( \cos{i} / \tan{\psi} \right)}$ \\

\noindent
Applying this to the Stokes vector ${\bf F}_{\rm planet}$, we thus 
obtained the following expressions for parameters $Q_{\rm orbit}$ and 
$U_{\rm orbit}$, as defined with respect to the orbital plane: \\

\noindent
$Q_{\rm orbit} = \hspace*{0.2cm} \cos 2\beta \hspace*{0.1cm} Q_{\rm planet} + \sin 2\beta \hspace*{0.1cm} U_{\rm planet}$ \\

\noindent
$U_{\rm orbit} = -\sin 2\beta \hspace*{0.1cm} Q_{\rm planet} + \cos 2\beta \hspace*{0.1cm} U_{\rm planet}$ \\

\noindent
These equations hold both for circular and elliptical orbits, because the ellipticity
does not change the values of angle $\psi$, only the change of $\psi$ in time.

In case of a non--zero value of $\kappa$, and additional rotation over the 
angle between the orbital plane used above and the actual reference plane 
should be performed. In particular, a rotation from the orbital plane
and a reference plane that we'll refer to as the \emph{detector plane}, 
perpendicular to the direction toward the observer's north, would be
described by the following equations: \\

\noindent
\hspace*{0.15cm} 
$0^\circ \leq \kappa \leq 90^\circ$ :
$Q_{\rm detector} = \cos 2\kappa \hspace*{0.1cm} Q_{\rm orbit} + \sin 2\kappa \hspace*{0.1cm} U_{\rm orbit}$ \\

\hspace*{1.7cm}
$U_{\rm detector} = -\sin 2\kappa \hspace*{0.1cm} Q_{\rm orbit} + \cos 2\kappa \hspace*{0.1cm} U_{\rm orbit}$ \\

\noindent $-90^\circ \leq \kappa \leq 0^\circ$ : 
$Q_{\rm detector} = \cos 2\kappa \hspace*{0.1cm} Q_{\rm orbit} - \sin 2\kappa \hspace*{0.1cm} U_{\rm orbit}$ \\

\hspace*{1.7cm}
$U_{\rm detector} = \sin 2\kappa \hspace*{0.1cm} Q_{\rm orbit} + \cos 2\kappa \hspace*{0.1cm} U_{\rm orbit}$ \\

\end{appendix}

\end{document}